\documentclass{ws-ijmpa}
\usepackage[super,compress]{cite}
\begin{document}

\markboth{Xue-wen Liu and Shun Zhou} {Texture Zeros for Dirac
Neutrinos and Current Experimental Tests}

\catchline{}{}{}{}{}
%
%%%%%%%%%%%%%%%%%%%%%%%%%%%%%%%%%%%%%%%%%%%%%%%%%%%%%%%%%%%%%%%%%%%%

\title{Texture Zeros for Dirac Neutrinos and Current
Experimental Tests}

\author{Xue-wen Liu}

\address{Department of Modern Physics \\
University of Science and Technology of China \\
Hefei 230026, China\\
xuewliu@mail.ustc.edu.cn}

\author{Shun Zhou}

\address{Department of Theoretical Physics \\
KTH Royal Institute of Technology \\
106 91 Stockholm, Sweden\\
shunzhou@kth.se}

\maketitle

\begin{history}
\received{Day Month Year}
\revised{Day Month Year}
\end{history}

\begin{abstract}
The Daya Bay and RENO reactor neutrino experiments have revealed
that the smallest neutrino mixing angle is in fact relatively large,
i.e., $\theta^{}_{13} \approx 9^\circ$. Motivated by this exciting
progress, we perform a systematic study of the neutrino mass matrix
$M^{}_\nu$ with one or two texture zeros, in the assumption that
neutrinos are Dirac particles. Among fifteen possible patterns with
two texture zeros, only three turn out to be favored by current
neutrino oscillation data at the $3\sigma$ level. Although all the
six patterns with one texture zero are compatible with the
experimental data at the $3\sigma$ level, the parameter space of
each pattern is strictly constrained. Phenomenological implications
of $M^{}_\nu$ on the leptonic CP violation and neutrino mass
spectrum are explored, and the stability of texture zeros against
the radiative corrections is also discussed.

\keywords{Texture Zeros; Neutrino Masses and Flavor Mixing; CP
Violation.}
\end{abstract}

\ccode{PACS numbers: 14.60.Lm, 14.60.Pq}

\section{Introduction}
The solar, atmospheric, accelerator and reactor neutrino experiments
have provided us with compelling evidence that neutrinos are massive
particles and they can transform from one flavor to another
\cite{PDG}. The lepton flavor mixing can be described by a $3\times
3$ unitary matrix $U$, which is usually parameterized through three
flavor mixing angles $(\theta^{}_{12}, \theta^{}_{23},
\theta^{}_{13})$ and one CP-violating phase $\delta$. To be
explicit, we adopt the following parametrization
\begin{equation}
U = \left(\begin{matrix}c^{}_{12} c^{}_{13} & s^{}_{12} c^{}_{13} &
s^{}_{13} \cr -c^{}_{12} s^{}_{23} s^{}_{13} - s^{}_{12} c^{}_{23}
e^{-i\delta} & -s^{}_{12} s^{}_{23} s^{}_{13} + c^{}_{12} c^{}_{23}
e^{-i\delta} & s^{}_{23} c^{}_{13} \cr -c^{}_{12} c^{}_{23}
s^{}_{13} + s^{}_{12} s^{}_{23} e^{-i\delta} & -s^{}_{12} c^{}_{23}
s^{}_{13} - c^{}_{12} s^{}_{23} e^{-i\delta} & c^{}_{23} c^{}_{13}
\end{matrix} \right)  \; ,
%     (1)
\end{equation}
where $s^{}_{ij} \equiv \sin \theta^{}_{ij}$ and $c^{}_{ij} \equiv
\cos \theta^{}_{ij}$ (for $ij = 12$, $23$, $13$) have been defined.
If neutrinos are Majorana particles, two additional CP-violating
phases $(\rho, \sigma)$ have to be introduced to fully describe the
flavor mixing. Thanks to the elegant neutrino oscillation
experiments, two neutrino mass-squared differences $(\delta m^2,
\Delta m^2)$ and two flavor mixing angles $(\theta^{}_{12},
\theta^{}_{23})$ have been measured with a reasonably good
precision. More recently, the Daya Bay \cite{Daya} and RENO
\cite{Reno} collaborations have clearly observed the disappearance
of $\bar{\nu}^{}_e$ from nuclear reactors, and revealed that the
smallest mixing angle is relatively large, i.e., $\theta^{}_{13}
\approx 9^\circ$. This is really a great news to the long-baseline
neutrino oscillation experiments, which aim to pin down the sign of
$\Delta m^2$ and the magnitude of the CP-violating phase $\delta$.
In spite of the great progress made in neutrino physics, our
understanding of neutrino properties is far from complete. For
instance, the absolute scale of neutrino masses is not yet
determined and whether neutrinos are Dirac or Majorana particles
remains an open question.

Since a convincing flavor theory is lacking, the approach of texture
zeros has been suggested to study the flavor problem for a long time
\cite{weinberg,wilczek,fritzsch,fritzsch1,fritzsch2}. The texture
zeros of a fermion mass matrix dynamically mean that the
corresponding matrix elements are sufficiently suppressed in
comparison with their neighboring counterparts \cite{FN}, and they
can help us to establish some simple and testable relations between
flavor mixing angles and fermion mass ratios \cite{X00,X04,X0406}.
In fact, a great number of works have been devoted to confronting
the zero textures of neutrino mass matrix with the neutrino
oscillation data.
\cite{rev1,rev2,FGM,xing1,xing2,Guo,More1,More2,More3,More4,
More5,More6,More7,More8,More9,More10,More11,More12,More13,More14,More15,
More16,More17,More18,More19,More20,More21,More22,More23,More24,More25,
FXZ,FXZ1,FXZ2,Xing05,Xing051,Xing052,Xing053,Xing054,Rode} However,
most of them have assumed neutrinos to be Majorana particles,
because various seesaw mechanisms for neutrino mass generation lead
to light Majorana neutrinos.

If neutrinos are Dirac particles, they can acquire masses exactly in
the same way as quarks and charged leptons do in the standard model.
In this scenario, it seems quite difficult to explain why the
neutrino Yukawa couplings are twelve orders of magnitude smaller
than the top-quark Yukawa coupling. Although this mass hierarchy
problem has never been well understood even for charged fermions, it
has been shown that the highly-suppressed Yukawa couplings for Dirac
neutrinos can naturally be achieved in the models with extra spacial
dimensions \cite{ED,ED1} or through radiative mechanisms
\cite{radiative,radiative1,radiative2,radiative3,radiative4,radiative5,
radiative6,radiative7}. Conservatively speaking, the most important
motivation for considering Dirac neutrinos is the fact that no
experiments have already excluded such a possibility. In the present
work, we simply assume neutrinos to be Dirac particles and perform a
systematic study of the neutrino mass matrix $M^{}_\nu$ with one or
two texture zeros.

Without loss of generality, we can take the mass matrix $M^{}_\nu$
for Dirac neutrinos to be Hermitian by redefining the right-handed
neutrino fields.\footnote{This can be done for the mass matrices of
Dirac fermions in the standard model. See Appendix A for a brief
proof and further discussions.} As $M^{}_\nu$ is Hermitian, three
independent off-diagonal matrix elements are in general complex,
while three independent diagonal ones are real. If $n$ of them are
taken to be vanishing (i.e., $M^{}_\nu$ has $n$ independent texture
zeros), then we shall arrive at
\begin{eqnarray}
^6{\bf C}_n = \frac{6!}{n! \left (6 - n\right )!}
%     (2)
\end{eqnarray}
different textures. There are totally fifteen two-zero textures of
$M^{}_\nu$, which can be classified into six categories:
\begin{eqnarray}
{\bf A^{}_1}: ~~ \left(\begin{matrix} 0 & 0 & \triangle \cr 0 &
\times & \triangle \cr \triangle^* & \triangle^* & \times
\end{matrix} \right) \; , ~~~~ {\bf A^{}_2}: ~~
\left(\begin{matrix} 0 & \triangle & 0 \cr \triangle^* & \times &
\triangle \cr 0 & \triangle^* & \times \end{matrix} \right) \; ;
%     (3)
\end{eqnarray}
%and
\begin{eqnarray}
{\bf B^{}_1}: ~~ \left(\begin{matrix} \times & \triangle & 0 \cr
\triangle^* & 0 & \triangle \cr 0 & \triangle^* & \times
\end{matrix} \right) \; , && {\bf B^{}_2}: ~~ \left(\begin{matrix}
\times & 0 & \triangle \cr 0 & \times & \triangle \cr \triangle^* &
\triangle^* & 0 \end{matrix} \right) \; , ~~~~~
\nonumber \\
{\bf B^{}_3}: ~~ \left(\begin{matrix} \times & 0 & \triangle \cr 0 &
0 & \triangle \cr \triangle^* & \triangle^* & \times
\end{matrix}\right) \; , && {\bf B^{}_4}: ~~ \left(\begin{matrix}
\times & \triangle & 0 \cr \triangle^* & \times & \triangle \cr 0 &
\triangle^* & 0 \end{matrix} \right) \; ;
%     (4)
\end{eqnarray}
%and
\begin{eqnarray}
{\bf C}: ~~ \left(\begin{matrix} \times & \triangle & \triangle \cr
\triangle^* & 0 & \triangle \cr \triangle^* & \triangle^* & 0
\end{matrix} \right) \; ;
%     (5)
\end{eqnarray}
%and
\begin{eqnarray}
{\bf D^{}_1}: ~~ \left(\begin{matrix} \times & \triangle & \triangle
\cr \triangle^* & 0 & 0 \cr \triangle^* & 0 & \times \end{matrix}
\right) \; , ~~~~ {\bf D^{}_2}: ~~ \left(\begin{matrix} \times &
\triangle & \triangle \cr \triangle^* & \times & 0 \cr \triangle^* &
0 & 0 \end{matrix} \right) \; ;
%     (6)
\end{eqnarray}
%and
\begin{eqnarray}
{\bf E^{}_1}: ~~ \left(\begin{matrix} 0 & \triangle & \triangle \cr
\triangle^* & 0 & \triangle \cr \triangle^* & \triangle^* & \times
\end{matrix} \right) \; , ~~~ {\bf E^{}_2}: ~~ \left(\begin{matrix} 0 &
\triangle & \triangle \cr \triangle^* & \times & \triangle \cr
\triangle^* & \triangle^* & 0 \end{matrix} \right) \; , ~~~ {\bf
E^{}_3}: ~~ \left(\begin{matrix} 0 & \triangle & \triangle \cr
\triangle^* & \times & 0 \cr \triangle^* & 0 & \times \end{matrix}
\right) \; ;
%     (7)
\end{eqnarray}
and
\begin{eqnarray}
{\bf F^{}_1}: ~~ \left(\begin{matrix} \times & 0 & 0 \cr 0 & \times
& \triangle \cr 0 & \triangle^* & \times \end{matrix} \right) \; ,
~~~~ {\bf F^{}_2}: ~~ \left(\begin{matrix} \times & 0 & \triangle
\cr 0 & \times & 0 \cr \triangle^* & 0 & \times \end{matrix} \right)
\; , ~~~~ {\bf F^{}_3}: ~~ \left(\begin{matrix} \times & \triangle &
0 \cr \triangle^* & \times & 0 \cr 0 & 0 & \times \end{matrix}
\right) \; ,
%     (8)
\end{eqnarray}
in which each ``$\times$" stands for a nonzero and real matrix
element, while each ``$\triangle$" for a nonzero and complex one.
Note that this classification is similar to that for the two-zero
textures of Majorana neutrino mass matrix
\cite{FGM,xing1,xing2,Guo}, which are symmetric and complex rather
than Hermitian. Although the one-zero textures are in general less
predictive than the two-zero ones, we shall consider them for
completeness. Assuming one of six independent matrix elements to be
zero, one can find out that there are six one-zero textures:
\begin{eqnarray}
{\bf P^{}_1}: ~~ \left(\begin{matrix} 0 & \triangle & \triangle \cr
\triangle^* & \times & \triangle \cr \triangle^* & \triangle^* &
\times \end{matrix} \right) \; , ~~~~ {\bf P^{}_2}: ~~
\left(\begin{matrix} \times & \triangle & \triangle \cr \triangle^*
& 0 & \triangle \cr \triangle^* & \triangle^* & \times \end{matrix}
\right) \; , ~~~~ {\bf P^{}_3}: ~~ \left(\begin{matrix} \times &
\triangle & \triangle \cr \triangle^* & \times & \triangle \cr
\triangle^* & \triangle^* & 0 \end{matrix} \right) \; ;
%     (9)
\end{eqnarray}
and
\begin{eqnarray}
{\bf P^{}_4}: ~~ \left(\begin{matrix} \times & 0 & \triangle \cr 0 &
\times & \triangle \cr \triangle^* & \triangle^* & \times
\end{matrix} \right) \; , ~~~~ {\bf P^{}_5}: ~~
\left(\begin{matrix} \times & \triangle & 0 \cr \triangle^* & \times
& \triangle \cr 0 & \triangle^* & \times \end{matrix} \right) \; ,
~~~~ {\bf P^{}_6}: ~~ \left(\begin{matrix} \times & \triangle &
\triangle \cr \triangle^* & \times & 0 \cr \triangle^* & 0 & \times
\end{matrix} \right) \; ,
%     (10)
\end{eqnarray}
where the notations are the same as those in Eqs. (3)-(8). It is
straightforward to observe that three two-zero patterns ${\bf
F^{}_{1,2,3}}$ in Eq. (8) can be excluded, because they lead to just
one nonzero flavor mixing angle.

We aim to confront the remaining twelve two-zero textures and six
one-zero textures with the latest global-fit results of current
neutrino oscillation data done by Fogli {\it et al.} in Ref.
\citen{Fogli}.\footnote{The global-fit analysis of neutrino
oscillation experiments has also been done in Refs. \citen{Valle}
and \citen{Schwetz} however, all the results are completely
consistent with each other at the $3\sigma$ level. Therefore, our
discussions will not be affected when the different global-fit
results are used.} Only three two-zero textures, i.e., ${\bf
A^{}_1}$, ${\bf A}^{}_2$, and ${\bf C}$, are found to be compatible
with current experimental data at the $3\sigma$ level, so are all
the six one-zero patterns, i.e., ${\bf P^{}_{\it i}}$ (for $i = 1,
2, \cdots, 6$). In particular, most physical consequences of those
viable patterns with one or two texture zeros have been explored in
an analytical way, and the stability of texture zeros against
radiative corrections is also discussed. We establish the
relationship between the location of texture zeros and CP
conservation, and demonstrate that the two-zero patterns ${\bf
A^{}_{1,2}}$ and the one-zero patterns ${\bf P^{}_{4,5,6}}$ lead to
CP conservation in the lepton sector.

The remaining parts of this paper are organized as follows. In
section 2, we give some general remarks on the texture zeros for
Dirac neutrinos. We show that it is possible to fully determine the
neutrino mass spectrum and the CP-violating phase $\delta$ for all
two-zero textures, as well as for the one-zero textures ${\bf
P^{}_{4,5,6}}$. The relationship between the location of texture
zeros and CP violation is pointed out. The stability of texture
zeros for Dirac neutrino mass matrix against the renormalization
group running is considered. Section 3 is devoted to the analytical
and numerical analyses of the two-zero textures of $M^{}_\nu$, while
section 4 to the one-zero textures. Finally we summarize our
conclusions in section 5.

\section{General Remarks}

\subsection{Important relations}

In the flavor basis where the charged-lepton mass matrix $M^{}_l$ is
diagonal, the Dirac neutrino mass matrix $M^{}_\nu$ can be
reconstructed in terms of three neutrino masses $(m^{}_1, m^{}_2,
m^{}_3)$ and the flavor mixing matrix $U$. Namely,
\begin{equation}
M^{}_\nu = U \left(\begin{matrix}\lambda^{}_1 & 0 & 0 \cr 0 &
\lambda^{}_2 & 0 \cr 0 & 0 & \lambda^{}_3 \end{matrix} \right)
U^\dagger \; ,
%     (11)
\end{equation}
where $\lambda^{}_1 = \eta \cdot m^{}_1$, $\lambda^{}_2 = \chi \cdot
m^{}_2$ and $\lambda^{}_3 = m^{}_3$ with $\eta, \chi = \pm 1$. Note
that the three eigenvalues of a general $3\times 3$ Hermitian matrix
are real, but not necessarily positive, so we have chosen the signs
of the first two eigenvalues relative to the third one as $\eta$ and
$\chi$. The parametrization of $U$ through three flavor mixing
angles $(\theta^{}_{12}, \theta^{}_{23}, \theta^{}_{13})$ and one
CP-violating phase $\delta$ has already been given in Eq. (1).

Now we explore the general consequences of the texture zeros of
$M^{}_\nu$. If one element of $M^{}_\nu$ is vanishing, i.e.,
$\left(M^{}_\nu\right)^{}_{\alpha \beta} = 0$, then we can obtain
the corresponding constraint among the flavor mixing parameters
\begin{equation}
\eta \cdot m^{}_1 U^{}_{\alpha 1} U^*_{\beta 1} + \chi \cdot m^{}_2
U^{}_{\alpha 2} U^*_{\beta 2} + m^{}_3 U^{}_{\alpha 3} U^*_{\beta 3}
=0 \; .
%     (12)
\end{equation}
Note that Eq. (12) implies one constraint condition in the case of
$\alpha = \beta$ (e.g., the one-zero textures ${\bf P^{}_{1,2,3}}$),
but two constraint conditions in the case of $\alpha \neq \beta$
(e.g., the one-zero textures ${\bf P^{}_{4,5,6}}$). In the former
case, one can derive from Eq. (12) that either $\eta = \chi = -1$ or
$\eta \cdot \chi = -1$ must hold and
\begin{equation}
|U^{}_{\alpha 2}|^2 = \frac{1}{1 - \chi \zeta} - \frac{1 - \eta
\xi}{1 - \chi \zeta} \cdot |U^{}_{\alpha 1}|^2 \; ,
%     (13)
\end{equation}
where $\xi \equiv m^{}_1/m^{}_3$ and $\zeta \equiv m^{}_2/m^{}_3$
have been defined. In the latter case, we can get two constraint
conditions by requiring both real and imaginary parts of the
left-hand side of Eq. (12) to be zero. More explicitly, we have
\begin{equation}
(\eta \xi - \chi \zeta) \cdot {\rm Im}\left[K^{\alpha
\beta}_{23}\right] = 0 \; ,
%     (14)
\end{equation}
and
\begin{equation}
{\rm Re}\left[K^{\alpha \beta}_{23}\right] = - \frac{1 - \eta \xi}{1
- \chi \zeta} \cdot {\rm Re}\left[K^{\alpha \beta}_{13}\right] \; ,
%     (15)
\end{equation}
where $K^{\alpha \beta}_{ij} \equiv U^{}_{\alpha i} U^*_{\alpha j}
U^*_{\beta i} U^{}_{\beta j}$. As will be shown later, the relations
in Eqs. (13), (14) and (15) are very useful in the determination of
flavor mixing parameters when we discuss the one-zero textures in
section 4.

If two independent elements of $M^{}_\nu$ are vanishing, i.e.,
$(M^{}_\nu)^{}_{ab} = (M^{}_\nu)^{}_{\alpha \beta} = 0$ with $ab
\neq \alpha \beta$ as shown in Eqs. (3)-(8), we can obtain
\begin{eqnarray}
\xi \equiv \frac{m^{}_1}{m^{}_3} &=&  \eta \cdot \frac{U^{}_{a3}
U^*_{b3} U^{}_{\alpha 2} U^*_{\beta 2} - U^{}_{a2} U^*_{b2}
U^{}_{\alpha 3} U^*_{\beta 3}}{U^{}_{a2} U^*_{b2} U^{}_{\alpha 1}
U^*_{\beta 1} - U^{}_{a1} U^*_{b1} U^{}_{\alpha 2} U^*_{\beta 2}}
\;, \nonumber \\
\zeta \equiv \frac{m^{}_2}{m^{}_3} &=& \chi \cdot \frac{U^{}_{a1}
U^*_{b1} U^{}_{\alpha 3} U^*_{\beta 3} - U^{}_{a3} U^*_{b3}
U^{}_{\alpha 1} U^*_{\beta 1}}{U^{}_{a2} U^*_{b2} U^{}_{\alpha 1}
U^*_{\beta 1} - U^{}_{a1} U^*_{b1} U^{}_{\alpha 2} U^*_{\beta 2}}
\;.
%     (16)
\end{eqnarray}
Since both $\xi$ and $\zeta$ are by definition real and
non-negative, the imaginary parts of the quantities on the
right-hand side of Eq. (16) have to disappear. This requirement may
lead us to the determination of the CP-violating phase $\delta$, as
we shall show below.

\subsection{Texture zeros and CP violation}

Now that $\theta^{}_{13}$ has been measured to be relatively large,
the CP-violating effects are promising to be discovered in the
long-baseline neutrino oscillation experiments, if the CP-violating
phase $\delta$ turns out to be not extremely small or very close to
$\pi$. As is well known, the CP violation is characterized by the
Jarlskog invariant ${\cal J}$, which is defined as
\cite{Jarlskog,Jarlskog1}
\begin{equation}
{\rm Im}\left[K^{\alpha \beta}_{ij}\right] \equiv {\cal J} \cdot
\sum_\gamma \epsilon^{}_{\alpha \beta \gamma} \sum_k
\epsilon^{}_{ijk} \; ,
%     (17)
\end{equation}
where $\epsilon^{}_{\alpha \beta \gamma}$ and $\epsilon^{}_{ijk}$
denote the Levi-Civita symbol, and $K^{\alpha \beta}_{ij}$ has been
defined below Eq. (15). It is now straightforward to discuss the CP
violation for the zero textures. Some comments are in order:
\begin{itemize}
\item {\sl CP-violating one-zero textures} -- In the case of
$(M^{}_\nu)^{}_{\alpha \beta} = 0$ with $\alpha \neq \beta$, Eq.
(14) implies either $\eta \xi = \chi \zeta$ or  ${\rm Im}
\left[K^{\alpha \beta}_{23}\right] = \pm {\cal J} = 0$. Note that
the solar neutrino experiments have established $m^{}_2 > m^{}_1$,
or equivalently $\zeta > \xi$, so we are left with ${\cal J} = 0$.
Therefore, CP violation is only possible for the patterns ${\bf
P^{}_{1,2,3}}$, and we have $\delta = 0$ or $\pi$ for the other
one-zero patterns.

\item {\sl CP-violating two-zero textures} -- Note that Eqs. (13),
(14) and (15) apply as well to the two-zero textures. Hence CP
violation is only possible for the textures with both $a = b$ and
$\alpha = \beta$. In other words, the patterns ${\bf C}$, ${\bf
E^{}_1}$, ${\bf E^{}_2}$ can lead to leptonic CP violation, while
$\delta = 0$ or $\pi$ holds for all the other two-zero textures.
\end{itemize}

In order to demonstrate the above observation of CP-violating
two-zero textures, we can directly calculate the imaginary part of
the first identity in Eq. (16). It turns out that
\begin{equation}
{\rm Re}\left[K^{ab}_{32}\right] \cdot {\rm Im}\left[K^{\alpha
\beta}_{21}\right] + {\rm Im}\left[K^{ab}_{32}\right] \cdot {\rm
Re}\left[K^{\alpha \beta}_{21}\right] - |U^{}_{a2}|^2 |U^{}_{b2}|^2
\cdot {\rm Im}\left[K^{\alpha \beta}_{31}\right] + (ab
\leftrightarrow \alpha \beta) = 0 \; ,
%     (18)
\end{equation}
where ``$ab \leftrightarrow \alpha\beta$" stands for the foregoing
terms with the exchange of $a \leftrightarrow \alpha$ and $b
\leftrightarrow \beta$ both in the superscripts and in the
subscripts. Since the patterns ${\bf F^{}_{1,2,3}}$ have already
been excluded, we need to consider only two different cases: (1) $a
= b$ and $\alpha = \beta$; (2) $a = b$ and $\alpha \neq \beta$. In
the first case, it is easy to see that the quantities on the
right-hand side of Eq. (16) are automatically real, so there is no
constraint on the CP-violating phase. In the second case, Eq. (18)
reduces to ${\rm Im}\left[K^{\alpha \beta}_{31}\right] = \pm {\cal
J} = 0$, which is consistent with our previous observation. Note
that the result for the second identity in Eq. (16) can be obtained
by exchanging the subscripts ``1" and ``2" in Eq. (18), however, it
doesn't give any new constraints. Hence we can conclude that only
one off-diagonal texture zero in the Hermitian Dirac neutrino mass
matrix is enough to ensure CP conservation in the lepton sector.

\subsection{Parameter counting}

Since we have assumed massive neutrinos to be Dirac particles, there
are seven physical parameters: three neutrino masses $( m^{}_1,
m^{}_2, m^{}_3)$, three flavor mixing angles $(\theta^{}_{12},
\theta^{}_{23}, \theta^{}_{13})$, and one CP-violating phase
$\delta$. The number of constraint relations caused by the texture
zeros depends on the location of zeros, so we count the parameters
for four distinct situations:
\begin{enumerate}
\item {\sl One off-diagonal zero} -- This category contains the
patterns ${\bf P^{}_{4,5,6}}$. As shown in Eqs. (14) and (15), there
are two independent constraint relations. Therefore, with the help
of current experimental measurements of three flavor mixing angles
$(\theta^{}_{12}, \theta^{}_{23}, \theta^{}_{13})$ and two neutrino
mass-squared differences, defined as \cite{Fogli}
\begin{eqnarray}
\delta m^2 \equiv m^2_2 - m^2_1 \; , ~~~~ \Delta m^2 = m^2_3 -
\frac{1}{2} \left(m^2_1 + m^2_2\right) \; ,
%     (19)
\end{eqnarray}
we can fully determine the neutrino mass spectrum and the
CP-violating phase. As shown in section 2.2, these patterns predict
CP conservation and thus $\delta = 0$ or $\pi$. Using Eq. (15) and
the following relation
\begin{equation}
R^{}_\nu \equiv \frac{\delta m^2}{|\Delta m^2|} = \frac{2(\zeta^2 -
\xi^2)}{|2 - (\zeta^2 + \xi^2)|} \; ,
%     (20)
\end{equation}
we can fix $\xi$ and $\zeta$. Thus three neutrino masses are given
by
\begin{equation}
m^{}_3 = \frac{\sqrt{\delta m^2}}{\sqrt{\zeta^2 - \xi^2}} \;, ~~~~
m^{}_2 = m^{}_3 \zeta \;, ~~~~ m^{}_1 = m^{}_3 \xi \; .
%     (21)
\end{equation}
The detailed discussions of ${\bf P^{}_{4,5,6}}$ will be given in
section 4.

\item {\sl One diagonal zero} -- This category contains the
patterns ${\bf P^{}_{1,2,3}}$. The texture zero leads to one
constraint condition in Eq. (13). Given the five experimental
observables, one parameter is left free for these patterns. If the
CP-violating phase is fixed, we can pin down the neutrino mass
spectrum by using Eqs. (13) and (20). On the other hand, if the
lightest neutrino mass is assumed, one can determine the
CP-violating phase.

\item {\sl One diagonal zero and one off-diagonal zero} -- This
category contains the patterns ${\bf A^{}_{1,2}}$, ${\bf
B^{}_{1,2,3,4}}$, ${\bf D^{}_{1,2}}$, and ${\bf E^{}_3}$. These two
texture zeros impose three constraint conditions, so both three
neutrino masses and the CP-violating phase can be calculated by
using experimental observables. Since these patterns are CP
conserving, we have $\delta = 0$ or $\pi$. Therefore, $\xi$ and
$\zeta$ can be calculated from Eq. (16), and the neutrino mass
spectrum is then given by Eq. (21). Note that $\xi$ and $\zeta$ have
to satisfy Eq. (20), so there will be one testable correlative
relation among the flavor mixing angles and neutrino masses.

\item {\sl Two diagonal zeros} -- These category contains the
patterns ${\bf C}$ and ${\bf E^{}_{1,2}}$. The texture zeros induce
two constraint relations, which together with five experimental
observables leads to the full determination of neutrino mass
spectrum and the CP-violating phase. This can be done as follows.
First, note that $\xi$ and $\zeta$ are functions of the CP-violating
phase $\delta$ as shown in Eq. (16). Then it is possible to
determine or constrain $\delta$ from Eq. (20). Once $\delta$ is
fixed, we can obtain $(\xi, \zeta)$ from Eq. (16), and thus neutrino
masses from Eq. (21). All the two-zero textures will be discussed in
great detail in section 3.
\end{enumerate}

Now we summarize the latest global-fit results of three flavor
mixing angles $(\theta^{}_{12}, \theta^{}_{23}, \theta^{}_{13})$ and
two neutrino mass-squared differences $(\delta m^2, \Delta m^2)$,
which will be taken as experimental observables to determine the
other flavor parameters. For the normal mass hierarchy with $\Delta
m^2 > 0$, it has been found at the $3\sigma$ level \cite{Fogli}
\begin{eqnarray}
0.259 \leq \sin^2 \theta^{}_{12} \leq 0.359 \;  ~~&{\rm or}&~~
30.6^\circ \leq \theta^{}_{12} \leq 36.8^\circ \; ,
\nonumber \\
0.331 \leq \sin^2 \theta^{}_{23} \leq 0.637 \;  ~~&{\rm or}&~~
35.1^\circ \leq \theta^{}_{23} \leq 53.0^\circ \; ,
\nonumber \\
0.017 \leq \sin^2 \theta^{}_{13} \leq 0.031 \; ~~&{\rm or}&~~
~~7.5^\circ \leq \theta^{}_{13} \leq 10.1^\circ \; ;
%     (22)
\end{eqnarray}
and
\begin{eqnarray}
6.99\times 10^{-5}~{\rm eV}^2 \leq &\delta m^2& \leq 8.18 \times
10^{-5}~{\rm eV}^2 \; ,
\nonumber \\
2.19\times 10^{-3}~{\rm eV}^2 \leq & \hspace{-0.25cm} + \Delta m^2
\hspace{-0.25cm} & \leq 2.62 \times 10^{-3}~{\rm eV}^2 \; .
%     (23)
\end{eqnarray}
For the inverted mass hierarchy with $\Delta m^2 < 0$, the global
analysis yields \cite{Fogli}
\begin{eqnarray}
0.259 \leq \sin^2 \theta^{}_{12} \leq 0.359 \;  ~~&{\rm or}&~~
30.6^\circ \leq \theta^{}_{12} \leq 36.8^\circ \; ,
\nonumber \\
0.335 \leq \sin^2 \theta^{}_{23} \leq 0.663 \;  ~~&{\rm or}&~~
35.4^\circ \leq \theta^{}_{23} \leq 54.5^\circ \; ,
\nonumber \\
0.017 \leq \sin^2 \theta^{}_{13} \leq 0.032 \; ~~&{\rm or}&~~
~~7.5^\circ \leq \theta^{}_{13} \leq 10.3^\circ \; ;
%     (24)
\end{eqnarray}
and
\begin{eqnarray}
6.99\times 10^{-5}~{\rm eV}^2 \leq &\delta m^2& \leq 8.18 \times
10^{-5}~{\rm eV}^2 \; ,
\nonumber \\
2.17\times 10^{-3}~{\rm eV}^2 \leq & \hspace{-0.25cm} - \Delta m^2
\hspace{-0.25cm} & \leq 2.61 \times 10^{-3}~{\rm eV}^2 \; ,
%     (25)
\end{eqnarray}
at the $3\sigma$ level. The best-fit values of three mixing angles
are $\theta^{}_{12} = 33.6^\circ$, $\theta^{}_{23} = 38.4^\circ$,
and $\theta^{}_{13} = 8.9^\circ$, while those of neutrino
mass-squared differences are $\delta m^2 = 7.54\times 10^{-5}~{\rm
eV}^2$ and $\Delta m^2 = 2.43\times 10^{-3}~{\rm eV}^2$. It is
interesting to note that the best-fit value of the CP-violating
phase is $\delta \sim \pi$, which happens to be consistent with the
prediction of the Dirac mass matrix with one off-diagonal texture
zero. However, there is no constraint on $\delta$ at the $3\sigma$
level.

\subsection{Stability of texture zeros}

The stability of texture zeros for Majorana neutrinos has already
been discussed in the literature \cite{RGEm,RGEm1,FXZ}, by using the
renormalization-group equations (RGEs) \cite{RGE,RGE1}. It has been
demonstrated that the texture zeros of Majorana neutrino mass matrix
are stable against the one-loop quantum corrections \cite{FXZ}. In
this subsection, we shall examine whether the stability of texture
zeros for Dirac neutrinos is maintained \cite{Rode}.

To accommodate Dirac neutrino masses, one can extend the standard
model with three right-handed neutrino singlets, and simply require
the lepton number conservation to forbid the Majorana neutrino mass
term. At the one-loop level, the RGEs for Dirac neutrinos and
charged leptons can be written as \cite{RGEd1, RGEd2, RGEd3, RGEd4,
RGEd5, book}
\begin{eqnarray}
16\pi^2 \frac{{\rm d} Y^{}_\nu}{{\rm d}t} &=& \left[ \alpha^{}_\nu +
C^\nu_\nu \left( Y^{}_\nu Y^\dagger_\nu \right) + C^l_\nu \left(
Y^{}_l Y^\dagger_l \right) \right] Y^{}_\nu \; , \nonumber \\
16\pi^2 \frac{{\rm d} Y^{}_l}{{\rm d}t} &=& \left[ \alpha^{}_l +
C^\nu_l \left( Y^{}_\nu Y^\dagger_\nu \right) + C^l_l \left( Y^{}_l
Y^\dagger_l \right) \right] Y^{}_l \; ,
%     (26)
\end{eqnarray}
where $t \equiv \ln(\mu/M^{}_Z)$ with $\mu$ being an arbitrary
renormalization scale, and $M^{}_Z$ the $Z$-boson mass. Here
$Y^{}_{\nu, l} = \sqrt{2} M^{}_{\nu, l}/v$ denotes respectively the
neutrino and charged-lepton Yukawa coupling matrices, and $v \approx
246~{\rm GeV}$ is the vacuum expectation value of the Higgs field.

In the standard model, the relevant coefficients are $C^\nu_\nu =
C^l_l = +3/2$, $C^l_\nu = C^\nu_l = -3/2$, $\alpha^{}_\nu =
-9g^2_1/20 - 9g^2_2/4 + T$ and $\alpha^{}_l = -9g^2_1/4 - 9g^2_2/4 +
T$, where $g^{}_1$ and $g^{}_2$ are gauge couplings, and $T = {\rm
Tr}\left[3\left( Y^{}_{\rm u} Y^\dagger_{\rm u} \right) + 3 \left(
Y^{}_{\rm d} Y^\dagger_{\rm d} \right) + \left( Y^{}_\nu
Y^\dagger_\nu \right) +\left( Y^{}_l Y^\dagger_l \right)\right]$
with $Y^{}_{\rm q}$ (for ${\rm q} = {\rm u}, {\rm d}$) being the
up-type and down-type quark Yukawa coupling matrices. The term
$Y^{}_\nu Y^\dagger_\nu$ in Eq. (26) can be safely neglected,
because neutrino masses are much smaller than charged-lepton masses.
In the flavor basis where the charged-lepton Yukawa matrix is
diagonal $Y^{}_l = {\rm Diag}\{y^{}_e, y^{}_\mu, y^{}_\tau\}$, one
can observe from the second identity in Eq. (26) that the
charged-lepton Yukawa coupling matrix remains diagonal as the energy
scale evolves. After solving the RGE for $Y^{}_\nu$, one can find
that the neutrino mass matrix at the weak scale $M^{}_Z$ is related
to that at a high-energy scale $\Lambda$ in the following way
\begin{equation}
M^{}_\nu(M^{}_Z) = I^{}_0 \left(\begin{matrix}I^{}_e & 0 & 0 \cr 0 &
I^{}_\mu & 0 \cr 0 & 0 & I^{}_\tau\end{matrix}\right)
M^{}_\nu(\Lambda) \; ,
%     (27)
\end{equation}
where the RGE evolution function $I^{}_0$ represents the overall
contribution from gauge and quark Yukawa couplings, while
$I^{}_\alpha$ (for $\alpha = e, \mu, \tau$) stand for the
contributions from charged-lepton Yukawa couplings. Because of
$m^{}_e \ll m^{}_\mu \ll m^{}_\tau$, we have $I^{}_e < I^{}_\mu <
I^{}_\tau$ and they will modify the structure of $M^{}_\nu$. In
contrast, $I^{}_0 \neq 1$ just affects the absolute scale of
neutrino masses. However, the texture zeros of $M^{}_\nu$ are stable
against the one-loop RGE corrections. Taking the pattern ${\bf
A^{}_1}$ for example, we have
\begin{equation}
M^{\bf A^{}_1}_\nu(\Lambda) = \left(\begin{matrix} 0 & 0 & a \cr 0 &
b & c \cr a^* & c^* & d\end{matrix}\right)
%     (28)
\end{equation}
at $\Lambda$, and thus
\begin{equation}
M^{\bf A^{}_1}_\nu(M^{}_Z) = I^{}_0 \left(\begin{matrix}0 & 0 & a
I^{}_e \cr 0 & b I^{}_\mu & c I^{}_\mu \cr a^* I^{}_\tau & c^*
I^{}_\tau & d I^{}_\tau\end{matrix}\right)
%     (29)
\end{equation}
at $M^{}_Z$. Note that the neutrino matrix $M^{\bf
A^{}_1}_\nu(M^{}_Z)$ is no longer exactly Hermitian. Nevertheless,
it can be shown that $I^{}_\alpha \approx 1$ (for $\alpha = e, \mu,
\tau$) hold as an excellent approximation in the standard model. As
a consequence, the previous important relations derived from texture
zeros are formally valid both at $\Lambda$ and $M^{}_Z$, at least in
the lowest-order approximation.

\section{Two-zero Textures}

\subsection{Analytical approximations}

First of all, we consider the two-zero textures and explore their
phenomenological implications in an analytical way. It has been
demonstrated that a permutation symmetry between the viable two-zero
patterns ${\bf A^{}_1}$ and ${\bf A^{}_2}$, ${\bf B^{}_1}$ and ${\bf
B^{}_2}$, or ${\bf B^{}_3}$ and ${\bf B^{}_4}$ for Majorana
neutrinos \cite{FXZ}. The existence of such a symmetry originates
from the fact that the location of texture zeros in each pair is
related by an exchange between the last two rows and columns of
$M^{}_\nu$. For the same reason, it is only necessary to study the
patterns ${\bf A^{}_1}$, $\bf B^{}_1$, $\bf B^{}_3$, $\bf D^{}_1$,
$\bf E^{}_1$ and $\bf E^{}_3$ of Dirac neutrino mass matrix, and the
results for the other patterns can be obtained with the replacements
$\theta^{}_{23} \to \pi/2- \theta^{}_{23}$ and $\delta \to \delta -
\pi$. Now we discuss these patterns by making some reasonable
approximations.

\begin{itemize}
\item {\bf Pattern $\bf A^{}_1$} with $(M^{}_\nu)^{}_{ee} =
(M^{}_\nu)^{}_{e\mu}=0$. As we have discussed, there is no CP
violation and thus $\delta = 0$ or $\pi$. In the case of $\delta =
0$, we obtain from Eq. (16) that
\begin{eqnarray}
\xi &=& +\eta \cdot \frac{s^{}_{13}}{c^2_{13}} \left(\frac{s^{}_{12}
s^{}_{23}}{c^{}_{12} c^{}_{23}} -s^{}_{13} \right) \;, \nonumber\\
\zeta &=& - \chi \cdot
\frac{s^{}_{13}}{c^2_{13}}\left(\frac{c^{}_{12} s^{}_{23}}{s^{}_{12}
c^{}_{23}} + s^{}_{13} \right)\;.
%  (30)
\end{eqnarray}
Obviously, $\eta = +1$ and $\chi = -1$ should be taken to ensure
that $\xi$ and $\zeta$ are non-negative. Since $s^2_{13} \ll 1$
still holds, Eq. (30) approximates to
\begin{eqnarray}
\xi &\approx& \sin \theta^{}_{13} \tan \theta^{}_{23} \tan
\theta^{}_{12}
\; , \nonumber \\
\zeta &\approx& \sin \theta^{}_{13} \tan \theta^{}_{23} \cot
\theta^{}_{12} \; .
%     (31)
\end{eqnarray}
Given $0.59 \leq \tan \theta^{}_{12} \leq 0.75$, $0.70 \leq \tan
\theta^{}_{23} \leq 1.3$, and $0.13 \leq \sin \theta^{}_{13} \leq
0.18$ at the $3\sigma$ level, Eq. (31) leads to $\xi < \zeta < 1$.
Hence only the normal neutrino mass hierarchy is allowed, i.e.,
$\Delta m^2 > 0$. Taking the best-fit values of three neutrino
mixing angles (i.e., $\theta^{}_{12} = 33.6^\circ$, $\theta^{}_{23}
= 38.4^\circ$, and $\theta^{}_{13} = 8.9^\circ$) and those of two
neutrino mass-squared differences (i.e., $\delta m^2 = 7.54\times
10^{-5}~{\rm eV}^2$ and $\Delta m^2 = 2.43\times 10^{-3}~{\rm
eV}^2$), one can figure out the neutrino mass spectrum
\begin{eqnarray}
m^{}_3 &\approx& \sqrt{\Delta m^2} = 4.9\times 10^{-2}~{\rm eV} \;,
\nonumber \\
m^{}_2 &\approx& m^{}_3 \sin \theta^{}_{13} \tan \theta^{}_{23} \cot
\theta^{}_{12} = 9.0 \times 10^{-3}~{\rm eV} \; , \nonumber \\
m^{}_1 &\approx& m^{}_3 \sin \theta^{}_{13} \tan \theta^{}_{23} \tan
\theta^{}_{12} = 4.0 \times 10^{-3}~{\rm eV} \; .
%     (32)
\end{eqnarray}
As shown in Eq. (20), there exists an interesting correlation
between three mixing angles and the neutrino mass ratios
\begin{equation}
R^{}_\nu = \frac{\delta m^2}{\Delta m^2} \approx \frac{4 \tan^2
\theta^{}_{23} \sin^2 \theta^{}_{13}}{\sin 2\theta^{}_{12} \tan
2\theta^{}_{12}} \; .
%     (33)
\end{equation}
Taking the best-fit values of three mixing angles, we can get
$R^{}_\nu \approx 0.027$ from Eq. (33). On the other hand, the
best-fit values of two neutrino mass-squared differences yield
$R^{}_\nu \approx 0.031$. Therefore,  ${\bf Pattern~ A^{}_1}$ is
well consistent with current oscillation data, and will be soon
tested in the future neutrino oscillation experiments. In the case
of $\delta = \pi$, we can obtain
\begin{eqnarray}
\xi &=& -\eta \cdot \frac{s^{}_{13}}{c^2_{13}} \left(\frac{s^{}_{12}
s^{}_{23}}{c^{}_{12} c^{}_{23}} + s^{}_{13} \right) \;, \nonumber\\
\zeta &=& + \chi \cdot
\frac{s^{}_{13}}{c^2_{13}}\left(\frac{c^{}_{12} s^{}_{23}}{s^{}_{12}
c^{}_{23}} - s^{}_{13} \right)\; ,
%  (34)
\end{eqnarray}
where $\eta = -1$ and $\chi = +1$ are implied. To the leading order
of $s^{}_{13}$, the phenomenological implications in the case of
$\delta = \pi$ are the same as those in the case of $\delta = 0$, so
we shall not discuss this case further. More precise measurements of
neutrino mixing angles are needed to distinguish between these two
cases.

\item ${\bf Pattern~ A^{}_2}$ with $(M^{}_\nu)^{}_{ee} =
(M^{}_\nu)^{}_{e\tau} =0$. All the analytical results of ${\bf
Pattern~ A^{}_2}$ can be obtained from those of ${\bf Pattern~
A^{}_1}$ with the replacements $\theta^{}_{23} \to
\pi/2-\theta^{}_{23}$ and $\delta \to \delta -\pi$. Therefore, in
the case of $\delta = 0$, we have
\begin{eqnarray}
\xi &=& -\eta \cdot \frac{s^{}_{13}}{c^2_{13}} \left(\frac{s^{}_{12}
c^{}_{23}}{c^{}_{12} s^{}_{23}} + s^{}_{13} \right) \;, \nonumber\\
\zeta &=& + \chi \cdot
\frac{s^{}_{13}}{c^2_{13}}\left(\frac{c^{}_{12} c^{}_{23}}{s^{}_{12}
s^{}_{23}} - s^{}_{13} \right)\; .
%  (35)
\end{eqnarray}
After taking $\eta = -1$ and $\chi = +1$ and neglecting the terms of
${\cal O}(s^2_{13})$, one can obtain
\begin{eqnarray}
\xi &\approx& \sin \theta^{}_{13} \cot \theta^{}_{23} \tan
\theta^{}_{12} \; , \nonumber \\
\zeta &\approx& \sin \theta^{}_{13} \cot \theta^{}_{23} \cot
\theta^{}_{12} \; .
%     (36)
\end{eqnarray}
Given $0.59 \leq \tan \theta^{}_{12} \leq 0.75$, $0.70 \leq \tan
\theta^{}_{23} \leq 1.3$, and $0.13 \leq \sin \theta^{}_{13} \leq
0.18$ at the $3\sigma$ level, we can verify that $\xi < \zeta < 1$,
implying that only the normal neutrino mass hierarchy is allowed.
The neutrino mass spectrum turns out to be
\begin{eqnarray}
m^{}_3 &\approx& \sqrt{\Delta m^2} \; , \nonumber \\
m^{}_2 &\approx& m^{}_3 \sin \theta^{}_{13} \cot \theta^{}_{23} \tan
\theta^{}_{12} \; , \nonumber \\
m^{}_1 &\approx& m^{}_3 \sin \theta^{}_{13} \cot \theta^{}_{23} \cot
\theta^{}_{12} \; .
%     (37)
\end{eqnarray}
As in the case of ${\bf Pattern~{A^{}_1}}$, there is a constraint
relation among three mixing angles and neutrino mass ratios. To the
leading order, we get
\begin{equation}
R^{}_\nu = \frac{\delta m^2}{\Delta m^2} \approx \frac{4 \cot^2
\theta^{}_{23} \sin^2 \theta^{}_{13}}{\sin 2\theta^{}_{12} \tan
2\theta^{}_{12}} \; .
%     (38)
\end{equation}
Although Eq. (38) is not fulfilled by the best-fit values of
$(\theta^{}_{12}, \theta^{}_{23}, \theta^{}_{13})$ and $(\delta m^2,
\Delta m^2)$, ${\bf Pattern~A^{}_2}$ is indeed compatible with
current oscillation data at the $3\sigma$ level, as will be
demonstrated by the numerical analysis in subsection 3.2. In a
similar way, one can discuss the case of $\delta = \pi$, for which
the analytical results can be obtained from Eqs. (30) and (31) by
replacing $\theta^{}_{23}$ with $\pi/2 - \theta^{}_{23}$.

\item {\bf Pattern $\bf B^{}_1$} with $(M^{}_\nu)^{}_{\mu\mu}
=(M^{}_\nu)^{}_{e\tau} = 0$. With the help of Eq. (16), we obtain
\begin{eqnarray}
\xi &=& \eta \cdot \frac{- s^{}_{12} c^{}_{12} s^3_{23} c^2_{13} \mp
(c^2_{12} c^2_{23} + s^2_{12} s^2_{23}) c^{}_{23} s^{}_{13} + 2
s^{}_{12} c^{}_{12} s^{}_{23} c^2_{23} s^2_{13}}{s^{}_{12} c^{}_{12}
s^{}_{23} c^2_{23} \pm (s^2_{12} - c^2_{12}) c^3_{23} s^{}_{13} +
s^{}_{12} c^{}_{12} s^{}_{23} s^2_{13} (1+c^2_{23})} \;,
\nonumber \\
\zeta &=& \chi \cdot \frac{- s^{}_{12} c^{}_{12} s^3_{23} c^2_{13}
\pm (c^2_{12} c^2_{23} + s^2_{12} s^2_{23}) c^{}_{23} s^{}_{13} + 2
s^{}_{12} c^{}_{12} s^{}_{23} c^2_{23} s^2_{13}}{s^{}_{12} c^{}_{12}
s^{}_{23} c^2_{23} \pm (s^2_{12} - c^2_{12}) c^3_{23} s^{}_{13} +
s^{}_{12} c^{}_{12} s^{}_{23} s^2_{13} (1+c^2_{23})} \;,
%     (39)
\end{eqnarray}
where the upper and lower signs refer to the cases of $\delta = 0$
and $\delta = \pi$, respectively. In the leading order
approximation, one can get $\eta = \chi = -1$ and $\xi \approx \zeta
\approx \tan^2 \theta^{}_{23}$. In the next-to-leading order
approximation, we find
\begin{equation}
\xi -\zeta \approx \pm \frac{4 \sin \theta^{}_{13}}{\sin
2\theta^{}_{12} \sin 2\theta^{}_{23}} \; .
%     (40)
\end{equation}
Since $\delta m^2 > 0$ or equivalently $\zeta > \xi$, only $\delta =
\pi$ is allowed. In this case, the constraint relation turns out to
be
\begin{eqnarray}
R^{}_\nu \approx \frac{2\sin \theta^{}_{13}}{\sin 2\theta^{}_{12}}
|\tan 2\theta^{}_{23}| \; .
%     (41)
\end{eqnarray}
Note that $R^{}_\nu \propto \sin \theta^{}_{13}$ in Eq. (41) may be
one order of magnitude larger than $R^{}_\nu \propto \sin^2
\theta^{}_{13}$ in Eq. (33) or Eq. (38). Taking the values of
$(\theta^{}_{12}, \theta^{}_{23}, \theta^{}_{13})$ in the $3\sigma$
ranges, one verify that $R^{}_\nu > 0.8$, which is obviously in
conflict with the experimental observation $R^{}_\nu < 0.038$.
Therefore, we conclude that ${\bf Pattern~B^{}_1}$ has already been
excluded by current neutrino oscillation data. Due to the
permutation symmetry between ${\bf Pattern~B^{}_1}$ and ${\bf
Pattern~B^{}_2}$, Eq. (41) is also applicable to the latter,
implying that ${\bf Pattern~B^{}_2}$ has been experimentally ruled
out as well.

\item {\bf Pattern $\bf B^{}_3$} with $(M^{}_\nu)^{}_{\mu\mu}
= (M^{}_\nu)^{}_{e\mu} = 0$. With the help of Eq. (16), we obtain
\begin{eqnarray}
\xi &=& -\eta \cdot \frac{s^{}_{23}}{c^{}_{23}} \cdot
\frac{s^{}_{12} s^{}_{23} \mp c^{}_{12} c^{}_{23}
s^{}_{13}}{s^{}_{12} c^{}_{23} \pm c^{}_{12} s^{}_{23} s^{}_{13}}
\;, \nonumber \\
\zeta &=& -\chi \cdot \frac{s^{}_{23}}{c^{}_{23}} \cdot
\frac{c^{}_{12} s^{}_{23} \pm s^{}_{12} c^{}_{23}
s^{}_{13}}{c^{}_{12} c^{}_{23} \mp s^{}_{12} s^{}_{23} s^{}_{13}}
\;,
%     (42)
\end{eqnarray}
where the upper and lower signs refer to the cases of $\delta = 0$
and $\delta = \pi$, respectively. In the leading order
approximation, $\eta = \chi = -1$ and $\xi \approx \zeta \approx
\tan^2 \theta^{}_{23}$ hold, as for ${\bf Pattern~B^{}_1}$. However,
in the next-to-leading order approximation, one gets
\begin{eqnarray}
\xi - \zeta \approx \mp \frac{4 \tan^2 \theta^{}_{23} \sin
\theta^{}_{13}}{\sin 2\theta^{}_{12} \sin 2\theta^{}_{23}} \; ,
%     (43)
\end{eqnarray}
and
\begin{eqnarray}
R^{}_\nu \approx \frac{2\sin^2 \theta^{}_{13}}{\sin 2\theta^{}_{12}}
\tan^2 \theta^{}_{23} |\tan 2\theta^{}_{23}| \; .
%     (44)
\end{eqnarray}
It is evident from Eq. (43) that only $\delta = 0$ is allowed,
because of $\zeta > \xi$. By taking the values of $(\theta^{}_{12},
\theta^{}_{23}, \theta^{}_{13})$ in the $3\sigma$ ranges, we can
obtain $R^{}_\nu > 0.4$, which is far outside of the $3\sigma$ range
of $R^{}_\nu$. Hence ${\bf Pattern~B^{}_3}$ is not compatible with
current neutrino oscillation data, nor is ${\bf Pattern~B^{}_4}$ due
to the permutation symmetry.

\item {\bf Pattern $\bf C$} with $(M^{}_\nu)^{}_{\mu\mu} =
(M^{}_\nu)^{}_{\tau\tau} = 0$. With the help of Eq. (16), we obtain
\begin{eqnarray}
\xi &=& \eta \cdot \frac{c^{}_{12} c^2_{13}}{s^{}_{13}} \cdot
\frac{c^{}_{12} (c^2_{23} - s^2_{23}) - 2 s^{}_{12} s^{}_{23}
c^{}_{23} s^{}_{13} c^{}_\delta}{2 s^{}_{12} c^{}_{12} s^{}_{23}
c^{}_{23} c^{}_\delta (1 + s^2_{13}) -(c^2_{12} - s^2_{12})(c^2_{23}
- s^2_{23})s^{}_{13} } \; , \nonumber \\
\zeta &=& \chi \cdot \frac{s^{}_{12} c^2_{13}}{s^{}_{13}} \cdot
\frac{s^{}_{12} (s^2_{23} - c^2_{23}) - 2 c^{}_{12} s^{}_{23}
c^{}_{23} s^{}_{13} c^{}_\delta}{2 s^{}_{12}c^{}_{12} s^{}_{23}
c^{}_{23} c^{}_\delta (1 + s^2_{13}) - (c^2_{12} -
s^2_{12})(c^2_{23} - s^2_{23})s^{}_{13} } \; ,
%     (45)
\end{eqnarray}
where we have defined $c^{}_\delta = \cos \delta$. Note that the
leptonic CP violation is allowed in this case. Generally speaking,
it is inappropriate to expand Eq. (45) in terms of $s^{}_{13}$ and
ignore the higher-order terms, because $c^{}_\delta$ and $c^2_{23} -
s^2_{23}$ could be vanishingly small according to current neutrino
oscillation data. But the analytical approximations in some
interesting limits deserve further discussions:
\begin{enumerate}
\item $\delta = \pi/2$ and $\theta^{}_{23} \neq \pi/4$. Insetting
$\delta = \pi/2$ into Eq. (45), one arrives at $\xi = - \eta \cot^2
\theta^{}_{13} \cos^2 \theta^{}_{12}/\cos 2\theta^{}_{12}$ and
$\zeta = +\chi \cot^2 \theta^{}_{13} \sin^2 \theta^{}_{12}/\cos
2\theta^{}_{12}$, which are independent of $\theta^{}_{23}$. Since
$\cos^2 \theta^{}_{12} > \sin^2 \theta^{}_{12}$ holds, we are led to
$\eta = -1$, $\chi = +1$, and $\xi > \zeta$. This observation
indicates that the maximal CP-violating phase is not allowed if
$\theta^{}_{23} \neq \pi/4$.

\item $\delta \neq \pi/2$ and $\theta^{}_{23} = \pi/4$. Assuming
$\theta^{}_{23} = \pi/4$ in Eq. (45), we can get $\eta = \chi = -1$
and $\xi = \zeta = \cos^2 \theta^{}_{13}/(1+\sin^2 \theta^{}_{13}) <
1$, which only depends on $\theta^{}_{13}$. Hence the maximal mixing
$\theta^{}_{23} = \pi/4$ is not favored if $\delta \neq \pi/2$, and
a tiny deviation of $\theta^{}_{23}$ from $\pi/4$ is necessary to
break the degeneracy between $m^{}_1$ and $m^{}_2$.

\item $\delta = \pi/2$ and $\theta^{}_{23} = \pi/4$. In this case,
one can verify that $|U^{}_{\mu i}|^2 = |U^{}_{\tau i}|^2$ (for $i =
1, 2, 3$) hold, so Eq. (16) is not applicable. The equality
$(M^{}_\nu)_{\mu \mu} = (M^{}_\nu)^{}_{\tau \tau}$ implies that only
one constraint relation is obtained by requiring $(M^{}_\nu)_{\mu
\mu} = 0$. Assuming $\delta = \pi/2$ and $\theta^{}_{23} = \pi/4$,
we find
\begin{equation}
\xi (c^2_{12} s^2_{13}+ s^2_{12}) + \zeta (s^2_{12} s^2_{13} +
c^2_{12}) = c^2_{13}
%     (46)
\end{equation}
in the case of $\eta = \chi = -1$. If the higher-order terms of
${\cal O}(s^2_{13})$ are neglected, then one obtains $\xi = (1 -
\zeta \cdot c^2_{12})/s^2_{12}$, implying that neither $\zeta > \xi
> 1$ nor $1 > \zeta > \xi$ is allowed. Consequently, both $\xi$ and
$\zeta$ should be close to one, and the deviations from one are of
${\cal O}(s^2_{13})$. Based on this observation, we define $\xi
\equiv 1 - \Delta \xi$ and $\zeta \equiv 1 - \Delta \zeta$, and then
solve Eqs. (20) and (46) for $\Delta \xi$ and $\Delta \zeta$. It is
straightforward to get
\begin{eqnarray}
\Delta \xi &=& \frac{(2 + R^{}_\nu) \sin^2 \theta^{}_{13}}{1 -
R^{}_\nu \cos 2\theta^{}_{12}/2} \; , \nonumber \\
\Delta \zeta &=& \frac{(2 - R^{}_\nu) \sin^2 \theta^{}_{13}}{1 -
R^{}_\nu \cos 2\theta^{}_{12}/2} \; .
%     (47)
\end{eqnarray}
Hence the neutrino masses are nearly degenerate, while both $\delta
\approx \pi/2$ and $\theta^{}_{23} \approx \pi/4$ are allowed and
expected to be valid simultaneously.

\item $\delta \neq \pi/2$ and $\theta^{}_{23} \neq \pi/4$. Furthermore,
we assume $c^{}_\delta \gg s^{}_{13}$ and then obtain
\begin{eqnarray}
\xi &\approx& -\eta \cdot \left(1 - \frac{\cot \theta^{}_{12} \cot
2\theta^{}_{23}}{\sin \theta^{}_{13} \cos \delta}\right)\; , \nonumber \\
\zeta &\approx& -\chi \cdot \left(1 + \frac{\tan \theta^{}_{12} \cot
2\theta^{}_{23}}{\sin \theta^{}_{13} \cos \delta}\right) \; .
%     (48)
\end{eqnarray}
If $\cot 2\theta^{}_{23} \cos \delta > 0$ holds, one can set $\chi =
-1$ and then obtain $\zeta > 1$, indicating the inverted mass
hierarchy $m^{}_2 > m^{}_1 > m^{}_3$. In this case, $\xi$ is
required to be larger than one but smaller than $\zeta$, and this
can be achieved by setting $\eta = +1$ and $\cot \theta^{}_{12} \cot
2\theta^{}_{23} > 2\sin \theta^{}_{13} \cos \delta > 2\cot
2\theta^{}_{12} \cot 2\theta^{}_{23}$. It is straightforward to
verify that $\cot 2\theta^{}_{23} \cos \delta \leq 0$ contradicts
with the requirement $\zeta > \xi$. Therefore, only the inverted
mass hierarchy is allowed.
\end{enumerate}

From the above discussions, one can observe that there exists a
small parameter space around $\delta = \pi/2$ and $\theta^{}_{23} =
\pi/4$, where the normal mass hierarchy is allowed and neutrino
masses are in fact nearly degenerate. In the main part of parameter
space, only the inverted mass hierarchy is consistent with current
oscillation data. For the general case, the numerical analysis will
be done in subsection 3.2.

\item {\bf Pattern $\bf D^{}_1$} with $(M^{}_{\nu})^{}_{\mu\mu}
= (M^{}_\nu)^{}_{\mu\tau} = 0$. With the help of Eq. (16), we obtain
\begin{eqnarray}
\xi &\approx& \cot \theta^{}_{12} \tan \theta^{}_{23} \;, \nonumber \\
\zeta &\approx& \tan \theta^{}_{12} \tan \theta^{}_{23} \; ,
%  (49)
\end{eqnarray}
where $\eta = -1$ and $\chi = +1$ are taken, and the higher-order
terms ${\cal O}(s^2_{13})$ have been safely neglected. Together with
the oscillation data, Eq. (49) indicates $\zeta < \xi$, which is in
contradiction with the experimental observation $m^{}_2 > m^{}_1$.
Therefore, this pattern is not viable, nor is ${\bf Pattern~D^{}_2}$
due to the permutation symmetry.

\item {\bf Pattern $\bf E^{}_1$} with $(M^{}_{\nu})^{}_{ee}
= (M^{}_\nu)^{}_{\mu\mu} = 0$. With the help of Eq. (16), we obtain
\begin{eqnarray}
\xi &\approx& \sin^2 \theta^{}_{12} \tan^2 \theta^{}_{23}/\cos
2\theta^{}_{12} \; , \nonumber \\
\zeta &\approx& \cos^2\theta^{}_{12} \tan^2 \theta^{}_{23}/\cos
2\theta^{}_{12} \; ,
%   (50)
\end{eqnarray}
in the leading-order approximation, where we have set $\eta = +1$
and $\chi = -1$. Since the neutrino mass ratios have been determined
as in Eq. (50), one can get
\begin{equation}
R^{}_\nu \approx \frac{4\tan^4 \theta^{}_{23} \cos
2\theta^{}_{12}}{\left|\tan^4 \theta^{}_{23} (1+\cos^2
2\theta^{}_{12}) - 4\cos^2 2\theta^{}_{12}\right|} \; ,
%     (51)
\end{equation}
implying $R^{}_\nu > 2\cos 2\theta^{}_{12}$ or $R^{}_\nu > \tan^4
\theta^{}_{23}/\cos 2\theta^{}_{12}$. Taking the allowed values of
$\theta^{}_{23}$ and $\theta^{}_{12}$ in the $3\sigma$ ranges, we
can derive $R^{}_\nu > 0.56$ or $R^{}_\nu > 0.50$, respectively.
Therefore, ${\bf Pattern~E^{}_1}$ is not consistent with current
neutrino oscillation data, nor is ${\bf Pattern~E^{}_2}$ according
to the permutation symmetry.

\item {\bf Pattern $\bf E^{}_3$} with $(M^{}_{\nu})^{}_{ee} =
(M^{}_\nu)^{}_{\mu\tau} = 0$. With the help of Eq. (16), we obtain
\begin{eqnarray}
\xi &\approx& \sin^2 \theta^{}_{12}/\cos 2\theta^{}_{12} \; ,
\nonumber \\
\zeta &\approx& \cos^2 \theta^{}_{12}/\cos 2\theta^{}_{12} \; ,
%   (52)
\end{eqnarray}
in the leading-order approximation, where we have chosen $\eta = -1$
and $\chi = +1$. The ratio of neutrino mass-squared differences
turns out to be
\begin{equation}
R^{}_\nu \approx \frac{4 \cos 2\theta^{}_{12}}{1 - 3\cos^2
2\theta^{}_{12}} > 4 \cos 2\theta^{}_{12} \; ,
%     (53)
\end{equation}
which is obviously disfavored by current experimental data.
\end{itemize}

In summary, we conclude that only three two-zero patterns (i.e.,
$\bf A^{}_1$, $\bf A^{}_2$ and $\bf C$) are consistent with current
neutrino oscillation data. The detailed numerical analysis of these
three patterns will be performed in the following subsection.

\subsection{Numerical Analysis}

We have numerically confirmed that only three patterns ${\bf
A^{}_1}$, ${\bf A^{}_2}$ and ${\bf C}$ are viable, as obtained from
the above analytical approximations. Our strategy for numerical
calculations is as follows:
\begin{enumerate}
\item For each of the twelve patterns of $M^{}_\nu$ in Eqs. (3)-(7)
we generate a set of random numbers of $(\theta^{}_{12},
\theta^{}_{23}, \theta^{}_{13}, \delta m^2)$ lying in their
$3\sigma$ ranges, which are already given in Eqs. (22)-(25). For the
CP-conserving patterns, we consider both cases of $\delta = 0$ and
$\delta = \pi$. For the CP-violating patterns, $\delta$ is allowed
to vary in the range $[0,2\pi)$.

\item With the above generated numbers, we can calculate the
other physical parameters of $M^{}_\nu$. First of all, the neutrino
mass ratios $\xi$ and $\zeta$ can be determined through three mixing
angles and the CP-violating phase. To be consistent with the
experimental data, the following two conditions should be satisfied:
(a) $m^{}_2 > m^{}_1$ or equivalently $\zeta^2 > \xi^2$; (b) since
only the neutrino mass hierarchies $m^{}_2 > m^{}_1 > m^{}_3$ and
$m^{}_3 > m^{}_2 > m^{}_1$ are allowed, we further require $(\zeta^2
- 1) (\xi^2 - 1) > 0$. With the calculated neutrino mass ratios
$\xi$ and $\zeta$, one can figure out the absolute neutrino masses
$m^{}_i$ (for $i=1, 2, 3$) by inserting the randomly generated
$\delta m^2$ into Eq. (21). Finally we compute $\Delta m^2$ via
neutrino masses $m^{}_i$, the patterns can be regarded as viable if
the computed $\Delta m^2$ is consistent with the experimental values
in Eqs. (23) and (25).

\item From all the points satisfying the above consistency conditions,
we can figure out three neutrino mass eigenvalues $(m^{}_1, m^{}_2,
m^{}_3)$ via Eq. (21). For the CP-violating patterns, it is possible
to calculate the Jarlskog invariant ${\cal J} \equiv s^{}_{12}
c^{}_{12} s^{}_{23} c^{}_{23} s^{}_{13} c^2_{13} s^{}_\delta$. To
present numerical results, we show the allowed regions of three
neutrino mass eigenvalues $(m^{}_1, m^{}_2, m^{}_3)$, those of three
mixing angles $(\theta^{}_{12}, \theta^{}_{23}, \theta^{}_{13})$,
and those of $({\cal J}, \delta)$ if CP violation is allowed.
\end{enumerate}

Our numerical results are shown in Figs. 1-3. Some comments and
discussions are in order:
\begin{itemize}
\item{\bf Pattern $\bf A^{}_1$}--
Our numerical results show that the patterns ${\bf A^{}_1}$ and
${\bf A}^{}_2$ survive current experimental tests, which confirms
our analytical calculations. The allowed ranges of neutrino mass
eigenvalues and mixing angles for ${\bf Pattern~A^{}_1}$ and ${\bf
Pattern~A^{}_2}$ are depicted in Fig. 1 and Fig. 2, respectively.
Since the patterns ${\bf A^{}_1}$ and ${\bf A^{}_2}$ are similar to
each other due to the permutation symmetry, here we focus on the
former. From Fig. 1, some interesting observations should be noted:
(a) Only the normal mass hierarchy $m^{}_1 < m^{}_2 < m^{}_3$ is
allowed. (b) The deviation of $\theta^{}_{23}$ from $\pi/4$ is
expected in both cases of $\delta = 0$ and $\delta = \pi$. However,
$\theta^{}_{23} < 45^\circ$ is favored for $\delta = 0$, while
$\theta^{}_{23} > 45^\circ$ for $\delta = \pi$. (c) The parameter
space in the case of $\delta = 0$ receives more stringent
constraints than that in the case of $\delta = \pi$. Therefore, more
precise measurements of neutrino mixing angles are required to
distinguish between the cases of $\delta = 0$ and $\delta = \pi$, as
well as between ${\bf Pattern~A^{}_1}$ and ${\bf Pattern~A^{}_2}$.

\item{\bf Pattern $\bf C$}--
Unlike the previous two patterns, the leptonic CP violation is
allowed for this pattern. Our numerical results are given in Fig. 3,
where only the inverted mass hierarchy has been considered. From the
analytical analysis in section 3.1, the normal mass hierarchy is
indeed allowed, if $\theta^{}_{23} \approx \pi/4$ and $\delta
\approx \pi/2$. However, the main parameter space points to the
inverted neutrino mass hierarchy. Although the mixing angles for
this pattern are not strictly constrained, the discovery of CP
violation in the future long-baseline neutrino oscillation
experiments will definitely single out ${\bf Pattern~C}$ among all
the two-zero textures as the true Dirac neutrino mass matrix.
\end{itemize}

Finally it is worth pointing out that we have performed a numerical
analysis of the two-zero textures of $M^{}_\nu$ by the oscillation
data at the $2\sigma$ level. We find that all these three patterns
are compatible with current experimental data at this level,
although the corresponding parameter space is somewhat smaller.

\section{One-zero Textures}

Now we proceed to consider the one-zero textures of Dirac neutrino
mass matrix $M^{}_\nu$. Since only one zero element is assumed,
these textures are not as predictive as the two-zero ones. On the
other hand, it will be no doubt that all the six patterns in Eqs.
(9) and (10) are compatible with current experimental data, which
has also been confirmed by numerical calculations. For simplicity,
we focus on the analytical analysis by making some reasonable
approximations.
\begin{itemize}
\item ${\bf Pattern~P^{}_1}$ with $(M^{}_\nu)^{}_{ee}=0$. With
the help of Eq. (12), we get
\begin{equation}
\eta \cdot \xi c^2_{12} c^2_{13} + \chi \cdot \zeta s^2_{12}
c^2_{13} + s^2_{13} = 0 \;.
%    (54)
\end{equation}
Since only one constraint relation exists, it is by no means
possible to pin down both the neutrino mass spectrum and the
CP-violating phase. From Eq. (54), one can observe that $\delta$ is
entirely unconstrained. In this case, however, neutrino mass
eigenvalues can be determined, as we shall show below. To find the
solutions to Eq. (54), one has to consider three different
possibilities: (1) $\eta = \chi = -1$; (2) $\eta = -1$ and $\chi =
+1$; (3) $\eta = +1$ and $\chi = -1$. Now we examine whether all
these possibilities are allowed by current experimental data.
\begin{enumerate}
\item If $\eta = \chi = -1$ is assumed, one can observe from Eq. (54)
that both $\xi$ and $\zeta$ should be of ${\cal O}(s^2_{13})$, and
furthermore obtain
\begin{equation}
\zeta = \tan^2 \theta^{}_{13}/\sin^2 \theta^{}_{12} - \xi \cot^2
\theta^{}_{12} \; ,
%     (55)
\end{equation}
from which $\xi < \tan^2 \theta^{}_{13}$ can be derived by requiring
$\zeta > \xi$. Since both $\xi$ and $\zeta$ are quite small in this
case, $R^{}_\nu \approx \zeta^2 - \xi^2$ holds as an excellent
approximation. To be explicit, one can find
\begin{eqnarray}
\xi &=& \frac{\tan^2 \theta^{}_{13} \cot^2 \theta^{}_{12} -
\sqrt{\tan^4 \theta^{}_{13} + R^{}_\nu \cos 2\theta^{}_{12}}
}{\cot^2 \theta^{}_{12} - 1} \; , \nonumber \\
\zeta &=& \frac{\tan^2 \theta^{}_{13} \tan^2 \theta^{}_{12} -
\sqrt{\tan^4 \theta^{}_{13} + R^{}_\nu \cos 2\theta^{}_{12}}
}{\tan^2 \theta^{}_{12} - 1} \; ,
%     (56)
\end{eqnarray}
which together with Eq. (21) leads to the neutrino mass spectrum. In
addition, the condition $\tan^2 \theta^{}_{13} \geq \sqrt{R^{}_\nu}
\sin^2 \theta^{}_{12}$ has to be fulfilled to guarantee a
non-negative $\xi$. Given $0.259 \leq \sin^2 \theta^{}_{12} \leq
0.359$, $0.017 \leq \tan^2 \theta^{}_{13} \leq 0.032$, and $0.027
\leq R^{}_\nu \leq 0.037$ at the $3\sigma$ level for the normal
neutrino mass hierarchy, it is straightforward to verify that such a
condition cannot be satisfied. Hence there is no solution in this
case.

\item If $\eta = +1$ and $\chi = -1$ are taken, then Eq. (54) can be
written as
\begin{equation}
\zeta = \tan^2 \theta^{}_{13}/\sin^2 \theta^{}_{12} + \xi \cot^2
\theta^{}_{12} \; .
%     (57)
\end{equation}
Different from the previous case, both $\zeta$ and $\xi$ need not to
be as small as $\tan^2 \theta^{}_{13}$. Requiring $\zeta > \xi$
leads to $\cot^2 \theta^{}_{12} > 1$, which is favored by current
oscillation data. Inserting Eq. (57) into Eq. (20), we can figure
out the neutrino mass ratios
\begin{eqnarray}
\xi &=& \frac{\tan^2 \theta^{}_{13} \cot^2 \theta^{}_{12} -
\sqrt{\tan^4 \theta^{}_{13} + R^{}_\nu \cos 2\theta^{}_{12}}}
{1 - \cot^2 \theta^{}_{12}} \; , \nonumber \\
\zeta &=& \frac{\tan^2 \theta^{}_{13} \tan^2 \theta^{}_{12} +
\sqrt{\tan^4 \theta^{}_{13} + R^{}_\nu \cos 2\theta^{}_{12}}} {1 -
\tan^2 \theta^{}_{12}} \; ,
%     (58)
\end{eqnarray}
where we have assumed $\xi^2 < \zeta^2 \ll 1$, and thus $R^{}_\nu
\approx \zeta^2 - \xi^2$. For the best-fit values $\theta^{}_{12} =
33.6^\circ$, $\theta^{}_{13} = 8.9^\circ$, and $R^{}_\nu = 0.031$,
the neutrino mass ratios are given as $\xi = 0.045$ and $\zeta =
0.22$, which justifies the assumption of $\xi^2 < \zeta^2 \ll 1$.
With the help of Eq. (21), the neutrino mass eigenvalues are found
to be $m^{}_3 \approx \sqrt{\delta m^2/(\zeta^2 - \xi^2)} =
0.04~{\rm eV}$, $m^{}_2 = \zeta m^{}_3 \approx 8.8\times
10^{-3}~{\rm eV}$ and $m^{}_1 = \xi m^{}_3 \approx 1.8\times
10^{-3}~{\rm eV}$.

\item If $\eta = -1$ and $\chi = +1$ are taken, then Eq. (54) turns
out to be
\begin{equation}
\zeta = \xi \cot^2 \theta^{}_{12} - \tan^2 \theta^{}_{13}/\sin^2
\theta^{}_{12} \; .
%     (59)
\end{equation}
The requirement $\zeta > \xi$ gives rise to a lower bound $\xi >
\tan^2 \theta^{}_{13}/\cos 2\theta^{}_{12}$. In a similar way to the
previous case, we assume $\xi^2 < \zeta \ll 1$ and thus $R^{}_\nu
\approx \zeta^2 - \xi^2$. Consequently, we arrive at
\begin{eqnarray}
\xi &=& \frac{\tan^2 \theta^{}_{13} +  \tan^2
\theta^{}_{12}\sqrt{\tan^4 \theta^{}_{13} + R^{}_\nu \cos
2\theta^{}_{12}}}{1 - \tan^2 \theta^{}_{12}} \; , \nonumber \\
\zeta &=& \frac{\tan^2 \theta^{}_{13} \tan^2 \theta^{}_{12} +
\sqrt{\tan^4 \theta^{}_{13} + R^{}_\nu \cos 2\theta^{}_{12}}} {1 -
\tan^2 \theta^{}_{12}} \; .
%     (60)
\end{eqnarray}
For the best-fit values $\theta^{}_{12} = 33.6^\circ$,
$\theta^{}_{13} = 8.9^\circ$, and $R^{}_\nu = 0.031$, the neutrino
mass ratios are given as $\xi = 0.13$ and $\zeta = 0.22$, which
justifies the assumption of $\xi^2 < \zeta^2 \ll 1$. The neutrino
masses turn out to be $m^{}_3 = \sqrt{\delta m^2/(\zeta^2 -\xi^2)}
\approx 0.049~{\rm eV}$, $m^{}_2 =\zeta m^{}_3 \approx 0.01~{\rm
eV}$, and $m^{}_1 =\xi m^{}_3 \approx 6.4\times 10^{-3}~{\rm eV}$.
\end{enumerate}

So we conclude that ${\bf Pattern~P^{}_1}$ is consistent with
current neutrino oscillation data, and only the normal mass
hierarchy is allowed. Additionally, the CP-violating phase $\delta$
is arbitrary.

\item ${\bf Pattern~P^{}_2}$ with $(M^{}_\nu)^{}_{\mu\mu}=0$.
With the help of Eq. (12), we get
\begin{equation}
\eta \cdot \xi (s^2_{12}  c^2_{23} + 2 {\cal J} t^{-1}_\delta
c^{-2}_{13} + c^2_{12} s^2_{23} s^2_{13}) + \chi \cdot \zeta
(c^2_{12} c^2_{23} - 2 {\cal J} t^{-1}_\delta c^{-2}_{13} + s^2_{12}
s^2_{23} s^2_{13}) + s^2_{23} c^2_{13} = 0 \; ,
%    (61)
\end{equation}
where ${\cal J} \equiv s^{}_{12} c^{}_{12} s^{}_{23} c^{}_{23}
s^{}_{13} c^2_{13} s^{}_\delta$ is the Jarlskog invariant in the
standard parametrization, and $t^{}_\delta \equiv \tan \delta$ has
been defined. For simplicity, we assume $\eta = \chi = -1$ and solve
Eq. (61) for $\xi$ and $\zeta$ in the leading-order approximation.
In this case, it is straightforward to observe $\xi \approx \zeta
\approx \tan^2 \theta^{}_{23}$. In order to figure out the
deviations of $\xi$ and $\zeta$ from $\tan^2 \theta^{}_{23}$, we
define $\xi \equiv \tan^2 \theta^{}_{23} - \Delta\xi$ and $\zeta
\equiv \tan^2 \theta^{}_{23} - \Delta \zeta$, and insert them back
into Eq. (61). Then one can see that both $\Delta \xi$ and $\Delta
\zeta$ should be of ${\cal O}(s^2_{13})$. Combining Eq. (61) with
Eq. (20), we obtain
\begin{eqnarray}
\Delta \xi &\approx& \frac{(1-\tan^4 \theta^{}_{23})R^{}_\nu +
2\tan^4 \theta^{}_{23} \sin^2 \theta^{}_{13}/(\cos^2 \theta^{}_{12}
\cos^2 \theta^{}_{23})}{2\tan^2 \theta^{}_{23}/\cos^2 \theta^{}_{12}
- \tan^2 \theta^{}_{23} (1 - \tan^2 \theta^{}_{12}) R^{}_\nu} \; ,
%     (62)
\end{eqnarray}
and $\Delta \zeta = \sin^2 \theta^{}_{13} \tan^2
\theta^{}_{23}/(\cos^2 \theta^{}_{12} \cos^2 \theta^{}_{23}) -
\Delta \xi \tan^2 \theta^{}_{12}$, where we have assumed the normal
mass hierarchy. Taking the best-fit values of $(\theta^{}_{12},
\theta^{}_{23},\theta^{}_{13})$ and $(\delta m^2, \Delta m^2)$ for
example, one can get $\xi = 0.59$ and $\zeta = 0.61$, and the
neutrino masses are $m^{}_3 = 6.4\times 10^{-2}~{\rm eV}$, $m^{}_2 =
3.9\times 10^{-2}~{\rm eV}$ and $m^{}_1 = 3.8\times 10^{-2}~{\rm
eV}$. Whether the neutrino mass hierarchy is normal or inverted
depends on the mixing angle $\theta^{}_{23}$. Similarly, one can
also analyze the case with $\eta \cdot \chi = -1$. It is worthwhile
to point out that $\delta$ is almost irrelevant to the determination
of neutrino masses, because its contribution to Eq. (61) is
suppressed by $\sin \theta^{}_{13}$. Therefore, $\delta$ is totally
arbitrary, as for ${\bf Pattern~P^{}_1}$.

Because of the permutation symmetry, the analytical results for
${\bf Pattern~P^{}_{3}}$ can be obtained by the replacements
$\theta^{}_{23} \to \pi/2 - \theta^{}_{23}$ and $\delta \to \pi -
\delta$. Thus we shall not discuss further about this case.

\item ${\bf Pattern~P^{}_4}$ with $(M^{}_\nu)^{}_{e\mu} = 0$. With the
help of Eq. (12), we get
\begin{equation}
-\eta \cdot \xi c^{}_{12} (s^{}_{12} c^{}_{23} \pm c^{}_{12}
s^{}_{23} s^{}_{13}) + \chi \cdot \zeta s^{}_{12} (c^{}_{12}
c^{}_{23} \mp s^{}_{12} s^{}_{23} s^{}_{13}) + s^{}_{13} s^{}_{23} =
0 \; ,
%     (63)
\end{equation}
where the upper and lower sign stands for $\delta = 0$ and $\delta =
\pi$, respectively. At the leading order, Eq. (63) approximates to
$(\eta \xi - \chi \zeta) = 2 \sin \theta^{}_{13} \tan
\theta^{}_{23}/\sin 2\theta^{}_{12}$, implying $\chi = -1$ and $\eta
= \pm1$. Take $\chi = -1$ and $\eta = +1$ for example. Since $\xi +
\zeta = 2 \sin \theta^{}_{13} \tan \theta^{}_{23}/\sin
2\theta^{}_{12}$, we see that $R^{}_\nu \approx \zeta^2 - \xi^2$
holds as an excellent approximation. Consequently, one gets
\begin{eqnarray}
\xi &=& \frac{4\tan^2 \theta^{}_{23} \sin^2 \theta^{}_{13} -
R^{}_\nu \sin^2 2\theta^{}_{12}}{4 \sin 2\theta^{}_{12} \tan
\theta^{}_{23} \sin \theta^{}_{13}} \;,\nonumber \\
\zeta &=& \frac{4\tan^2 \theta^{}_{23} \sin^2 \theta^{}_{13} +
R^{}_\nu \sin^2 2\theta^{}_{12}}{4 \sin 2\theta^{}_{12} \tan
\theta^{}_{23} \sin \theta^{}_{13}} \; ,
%     (64)
\end{eqnarray}
implying the normal neutrino mass hierarchy. Taking the best-fit
values of $(\theta^{}_{12}, \theta^{}_{23},\theta^{}_{13})$ and
$(\delta m^2, \Delta m^2)$, we have $\xi = 0.07$ and $\zeta = 0.19$,
and thus the neutrino masses $m^{}_3 = 4.9\times 10^{-2}~{\rm eV}$,
$m^{}_2 = 9.3\times 10^{-3}~{\rm eV}$ and $m^{}_1 = 3.4\times
10^{-3}~{\rm eV}$. In the case $\chi = -1$ and $\eta = -1$, it is
straightforward to verify that only the inverted mass hierarchy is
allowed.

Because of the permutation symmetry, the analytical results for
${\bf Pattern~P^{}_{5}}$ can be obtained by the replacements
$\theta^{}_{23} \to \pi/2 - \theta^{}_{23}$ and $\delta \to \pi -
\delta$.

\item {\bf Pattern $\bf P^{}_6$: $(M^{}_\nu)^{}_{\mu\tau}=0$}, we have
\begin{eqnarray}
\eta \xi [s^2_{12} \mp 2\cot 2\theta^{}_{23} c^{}_{12} s^{}_{12}
s^{}_{13} - c^2_{12} s^2_{13}]  + \chi \zeta [c^2_{12} \pm 2\cot
2\theta^{}_{23} c^{}_{12} s^{}_{12} s^{}_{13} - s^2_{12}  s^2_{13}]
= c^2_{13} \; ,
%    (65)
\end{eqnarray}

where the upper and lower sign stands for $\delta = 0$ and $\delta =
\pi$, respectively. In the leading-order approximation, we can
obtain $\eta \xi s^2_{12} + \chi \zeta c^2_{12} = c^2_{13}$. Note
that neutrino oscillation data indicate $c^2_{12} > s^2_{12}$ and
$\zeta > \xi$, so only two possibilities need to be considered: (1)
$\eta = \chi = +1$; and (2) $\eta = - 1$ and $\chi = +1$. Now we
examine whether these two possibilities are compatible with current
neutrino oscillation data:
\begin{enumerate}
\item If $\eta = \chi = +1$ is assumed, then $\xi \approx \zeta
\approx \cos^2 \theta^{}_{13}$ holds. In order to figure out the
deviations of $\xi$ and $\zeta$ from $\cos^2 \theta^{}_{13}$, we
define $\xi \equiv \cos^2 \theta^{}_{13} - \Delta \xi$ and $\zeta
\equiv \cos^2 \theta^{}_{13} - \Delta \zeta$, and insert them back
into Eq. (65). Then we obtain $\Delta \zeta = -\tan^2 \theta^{}_{12}
\Delta \xi$. Combing Eq. (65) with Eq. (20), we obtain $\Delta \xi
\approx R^{}_\nu \sin^2 \theta^{}_{13} \cos^2 \theta^{}_{12}$.
Taking the best-fit values of neutrino mixing parameters, we finally
get $m^{}_3 = 0.38~{\rm eV}$. Considering $m^{}_1 \approx m^{}_2
\approx m^{}_3$, we can see this might violate the cosmological
limit $\sum m^{}_i \equiv m^{}_1 + m^{}_2 + m^{}_3 < 0.61~{\rm eV}$
from the cosmic microwave background and the large-scale structure
observations \cite{bound}.

\item If $\eta = -1$ and $\chi = +1$ are taken,
we can get $\xi s^2_{12} + 1 \approx \zeta c^2_{12}$. Current
oscillation data indicate $s^2_{12} < c^2_{12}$, thus $\xi > 1$ is
required, implying the inverted neutrino mass hierarchy $m^{}_3 <
m^{}_1 < m^{}_2$. Furthermore, combining $\xi \approx \zeta \cot^2
\theta^{}_{12} - 1 / \sin^2 \theta^{}_{12} > 1$ with $R^{}_\nu
\approx 2 (\zeta^2 - \xi^2) / (\zeta^2 + \xi^2)$, we can obtain
$\zeta = 1 / (\cos^2 \theta^{}_{12} - \sin^2 \theta^{}_{12} \sqrt{(2
+ R^{}_\nu)/(2 - R^{}_\nu)})$. Taking the best-fit values of
neutrino mixing parameters, one can obtain $\zeta = 2.62$ and $\xi =
2.60$, and the neutrino masses are $m^{}_3 = 2.7\times 10^{-2}~{\rm
eV}$, $m^{}_2 = 7.1\times 10^{-2}~{\rm eV}$ and $m^{}_1 = 7.0\times
10^{-2}~{\rm eV}$.
\end{enumerate}

Therefore, we conclude that ${\bf Pattern~P^{}_6}$ is completely
consistent with current neutrino oscillation data, and only the
inverted mass hierarchy is allowed. However, the absolute neutrino
masses might exceed the cosmological bound in the case of a
nearly-degenerate neutrino mass spectrum.
\end{itemize}

\section{Summary}

In light of the recent measurements of $\theta^{}_{13}$ and the
latest global-fit analysis of neutrino oscillation experiments, we
have performed a systematic study of the Dirac neutrino mass matrix
$M^{}_\nu$ with two independent texture zeros or one texture zero.
It turns out that three two-zero patterns (i.e., ${\bf A^{}_{1,2}}$
and ${\bf C}$) and all six one-zero patterns(i.e., ${\bf
P^{}_{1,2,3,4,5,6}}$) can survive current experimental tests at the
$3\sigma$ level. In fact, all of them are found to be compatible
with the oscillation data even at the $2\sigma$ level. Analytical
analyses have been done for both two-zero and one-zero textures in
order to understand why they are favored or disfavored by current
experimental data. Moreover, the allowed parameter space for three
viable two-zero textures has been obtained through a detailed
numerical analysis. The following is a brief summary of our
conclusions:
\begin{itemize}
\item In the basis where the charged-lepton mass matrix is diagonal,
neutrino masses, flavor mixing angles, and CP-violating phase are
determined by the Dirac neutrino mass matrix $M^{}_\nu$, which can
be further made Hermitian by redefining the right-handed singlet
neutrino fields. We demonstrate that one vanishing off-diagonal
element in the Hermitian neutrino mass matrix is enough to guarantee
CP conservation in the lepton sector.

\item Among fifteen two-zero textures of $M^{}_\nu$, only three (i.e.,
${\bf A^{}_1}$, ${\bf A^{}_2}$ and ${\bf C}$) turn out to be
consistent with current neutrino oscillation data at the $3\sigma$
level. We have explained in detail why the other patterns are
disfavored. For ${\bf Pattern~A^{}_1}$ and ${\bf Pattern~A^{}_2}$,
only the normal neutrino mass hierarchy is allowed, and there is no
CP violation. For ${\bf Pattern~C}$, both normal and inverted mass
hierarchies are in principle allowed. However, the normal mass
hierarchy is possible only if $\delta \approx \pi/2$ and
$\theta^{}_{23} \approx \pi/4$. The precise measurements of neutrino
mixing angles and leptonic CP violation are needed to distinguish
among these currently viable patterns.

\item All the six one-zero textures are compatible with current
neutrino oscillation data. For the patterns ${\bf P^{}_{4,5,6}}$,
even only one texture zero is assumed, it is possible to fully
determine neutrino mass eigenvalues and the CP-violating phase in
terms of the observed three mixing angles $(\theta^{}_{12},
\theta^{}_{23}, \theta^{}_{13})$ and two neutrino mass-squared
differences $(\delta m^2, \Delta m^2)$. For the patterns ${\bf
P^{}_{1,2,3}}$, even if one free parameter remains, the full
determination of neutrino mass spectrum is possible, at least in the
first-order approximation. Although the parameter space for all
these patterns has received some constraints, it seems impossible to
exclude them experimentally in the near future.
\end{itemize}

It is worthwhile to stress that whether neutrinos are Dirac or
Majorana particles remains an open question. At present, the only
feasible way to demonstrate that neutrinos are Majorana particles is
to observe the neutrinoless double-beta decays \cite{review}. If
neutrinos are Dirac particles, such a lepton-number-violating
process is forbidden. Therefore, it really makes sense to consider
Dirac neutrinos, and to study the flavor mixing and CP violation in
this scenario. The ongoing and upcoming neutrino oscillation
experiments are expected to precisely measure the neutrino mixing
parameters, in particular the smallest mixing angle
$\theta^{}_{13}$, the deviation of $\theta^{}_{23}$ from $\pi/4$ and
the Dirac CP-violating phase $\delta$. The sensitivity of future
cosmological observations to the sum of neutrino masses $\sum
m^{}_i$ will probably reach $\sim 0.05~{\rm eV}$ in the near future.
We therefore expect that some patterns of the two-zero textures of
the neutrino mass matrix $M^{}_\nu$ for Dirac neutrinos might be
excluded or only marginally allowed by tomorrow's data, and those
surviving the experimental tests should shed light on the underlying
flavor structure of massive neutrinos.

\section*{Acknowledgments}

We are indebted to Prof. Zhi-zhong Xing for suggesting such an
investigation and helpful discussions. One of us (X.W.L.) would like
to thank Dr. Yu-feng Li for useful discussions, and Theoretical
Physics Division of IHEP for financial support and hospitality in
Beijing. The main part of this work was done at IHEP, Beijing
(X.W.L.) and at Max-Planck-Institut f\"{u}r Physik, M\"{u}nchen
(S.Z.). This work was partially supported by the European Union FP7
INVISIBLES (Marie Curie Actions, PITN-GA-2011-289442) and by the
G\"{o}ran Gustafsson Foundation.

\appendix

\section{Matrix Polar Decomposition}

In this appendix, we prove that the mass matrix $M^{}_\nu$ of Dirac
neutrinos can be made Hermitian by redefining the right-handed
singlet neutrino fields. In the extension of the standard model with
three right-handed singlet neutrinos $\nu^{}_{\rm R}$, the
Lagrangian relevant for lepton masses reads
\begin{equation}
-{\cal L}^{}_l = \overline{\ell^{}_{\rm L}} Y^{}_l l^{}_{\rm R} H +
\overline{\ell^{}_{\rm L}} Y^{}_\nu \nu^{}_{\rm R} \tilde{H} + {\rm
h.c.} \; ,
%     (A1)
\end{equation}
where $\ell^{}_{\rm L}$ and $\tilde{H} \equiv i\sigma^{}_2 H^*$
denote the lepton and Higgs doublets, $l^{}_{\rm R}$ stand for the
charged-lepton singlets, $Y^{}_l$ and $Y^{}_\nu$ are the $3\times 3$
Yukawa coupling matrices for charged leptons and neutrinos.

Now one can perform the basis transformation $\ell^\prime_{\rm L} =
U^\dagger_l \ell^{}_{\rm L}$ and $l^\prime_{\rm R} = V^\dagger_l
l^{}_{\rm R}$ to diagonalize the charged-lepton Yukawa coupling
matrix, i.e., $U^\dagger_l Y^{}_l V^{}_l = D^{}_l \equiv {\rm
Diag}\{y^{}_e, y^{}_\mu, y^{}_\tau\}$. In this basis, the neutrino
Yukawa coupling matrix turns out to be $Y^\prime_\nu = U^\dagger_l
Y^{}_\nu$. After the electroweak gauge symmetry breaking, the lepton
mass term is given by
\begin{equation}
-{\cal L}^{}_{\rm m} = \overline{l^{}_{\rm L}} M^{}_l l^{}_{\rm R} +
\overline{\nu^{}_{\rm L}} M^{}_\nu \nu^{}_{\rm R} + {\rm h.c.} \; ,
%     (A2)
\end{equation}
where $M^{}_l = D^{}_l v$ and $M^{}_\nu = Y^\prime_\nu v$ with $v
\approx 174~{\rm GeV}$ being the vacuum expectation value of the
Higgs field. In the chosen basis, $M^{}_l = {\rm Diag}\{m^{}_e,
m^{}_\mu, m^{}_\tau\}$ is diagonal with the charged-lepton masses
$m^{}_\alpha = y^{}_\alpha v$ (for $\alpha = e, \mu, \tau$), while
$M^{}_\nu$ is in general an arbitrary $3\times 3$ complex matrix. In
the following, we shall show that the neutrino mass matrix can be
decomposed as $M^{}_\nu = S^{}_\nu \cdot V^{}_\nu$, where $S^{}_\nu$
is an Hermitian matrix and $V^{}_\nu$ is a unitary matrix.
Therefore, we can redefine the right-handed neutrino field
$\nu^\prime_{\rm R} = V^{}_\nu \nu^{}_{\rm R}$ and thus the neutrino
mass matrix is Hermitian.

Finally, we prove that an $n\times n$ complex matrix $M$ can be
decomposed into $M = S \cdot V$, where $S$ is an $n\times n$
positive semi-definite Hermitian matrix and $V$ is an $n\times n$
unitary matrix. First, we define $H \equiv M M^\dagger$, which is
obviously an Hermitian matrix with non-negative eigenvalues, so it
can be diagonalized by a unitary transformation $U^\dagger H U = D^2
\equiv {\rm Diag}\{\lambda^2_1, \lambda^2_2, \cdots, \lambda^2_n\}$,
where $\lambda^{}_i \geq 0$ for $i = 1, 2, \cdots, n$. Then, we
choose $S = \sqrt{H} = \sqrt{M M^\dagger}$ with $\sqrt{H} \equiv U D
U^\dagger$, and it is straightforward to verify $S^\dagger = S$ and
$S^2 = H$. Note that $S$ is positive semi-definite and Hermitian,
and $M$, $H$ and $S$ have the same rank $m \leq n$. By definition,
the orthonormal eigenvectors $\psi^{}_i$ (for $i = 1, 2, \cdots, n$)
of $S$ are just the column vectors of the unitary matrix $U$, i.e.,
$U = (\psi^{}_1, \psi^{}_2, \cdots, \psi^{}_n)$ and $S \psi^{}_i =
\lambda^{}_i \psi^{}_i$. Furthermore, we introduce
\begin{equation}
\phi^{}_i = \left\{ \begin{array}{cc}
                         \displaystyle \frac{1}{\lambda^{}_i} M^\dagger
                         \psi^{}_i, & 1 \leq i \leq m \\
                         \varphi^{}_i, & m < i \leq n
                       \end{array}
\right. \; ,
\end{equation}
where $\{\varphi^{}_{m+1}, \varphi^{}_{m+2}, \cdots, \varphi^{}_n\}$
can be constructed from $\{\psi^{}_{m+1}, \psi^{}_{m+2}, \cdots,
\psi^{}_n\}$ such that $\{\phi^{}_i\}$ is a complete orthonormal
basis, i.e., $\phi^\dagger_i \phi^{}_j = \delta^{}_{ij}$. Defining
an $n\times n$ matrix $V$, whose element is given by $V =
(\psi^{}_1, \psi^{}_2, \cdots, \psi^{}_n) (\phi^{}_1, \phi^{}_2,
\cdots, \phi^{}_n)^\dagger$, we can prove that $M$ and $SV$ behave
in the same way on the basis vectors $\{\psi^{}_i\}$. The proof is
as follows:
\begin{enumerate}
\item Since either $\{\phi^{}_i\}$ or $\{\psi^{}_i\}$ is a complete
set of orthonormal vectors, we have two unitary matrices $U =
(\psi^{}_1, \psi^{}_2, \cdots, \psi^{}_n)$ and $U^\prime =
(\phi^{}_1, \phi^{}_2, \cdots, \phi^{}_n)$. Thus it is obvious that
$V = U U^{\prime \dagger}$ is a unitary matrix.

\item We show $\psi^\dagger_i SV = \psi^\dagger_i M$. More explicitly,
\begin{equation}
\psi^\dagger_i SV = \left\{ \begin{array}{cc}
                         \lambda^{}_i \psi^\dagger_i
                         V = \psi^\dagger_i M, & 1 \leq i \leq m \\
                         0, & m < i \leq n
                       \end{array}
\right. \; ,
\end{equation}
where we have used $\psi^\dagger_i M = \phi^\dagger_i M = 0$ for
$m+1 \leq i \leq n$. Since $\{\psi^{}_i\}$ is a complete set of
orthonormal basis, we arrive at $M = S \cdot V$.
\end{enumerate}

Although it is always possible to choose $S$ to be positive
semi-definite and Hermitian, we have considered the Dirac neutrino
mass matrices $M^{}_\nu$ to be more general in the sense that the
eigenvalues $\lambda_i$ can be either positive or negative.

%%%%%%%%%%%%%%%%%%%%% Fig. 1 %%%%%%%%%%%%%%%%%%%%%%%%%%%%%%
\begin{figure}
\centerline{\psfig{file=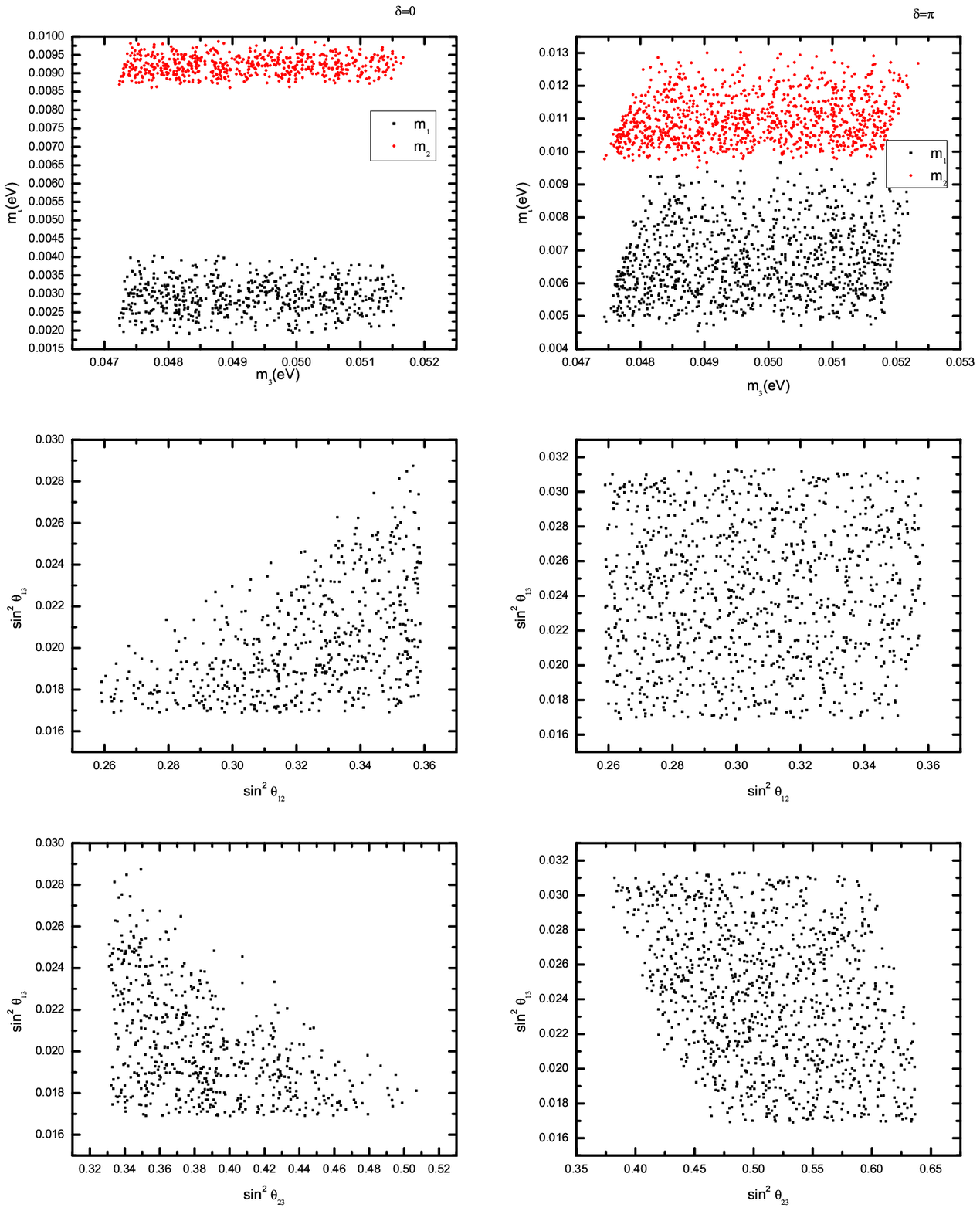,bbllx=3cm,bblly=13cm,bburx=18cm,bbury=28cm,%
width=13cm,height=13cm,angle=0,clip=0}} \vspace{3.5cm}
\caption{${\bf Pattern~A^{}_1}$ of $M^{}_\nu$: The allowed regions
of neutrino masses $(m^{}_1, m^{}_2, m^{}_3)$ and neutrino mixing
angles $(\sin^2 \theta^{}_{12}, \sin^2 \theta^{}_{23}, \sin^2
\theta^{}_{13})$, where the left column is for $\delta=0$ while the
right column for $\delta=\pi$. The global-fit data of two neutrino
mass-squared differences $(\delta m^2, \Delta m^2)$ and three
neutrino mixing angles $(\theta^{}_{12}, \theta^{}_{23},
\theta^{}_{13})$ at the $3\sigma$ level have been input.}
\end{figure}
%%%%%%%%%%%%%%%%%%%%%%%%%%%%%%%%%%%%%%%%%%%%

%%%%%%%%%%%%%%%%%%%%% Fig. 2 %%%%%%%%%%%%%%%%%%%%%%%%%%%%%%
\begin{figure}
\centerline{\psfig{file=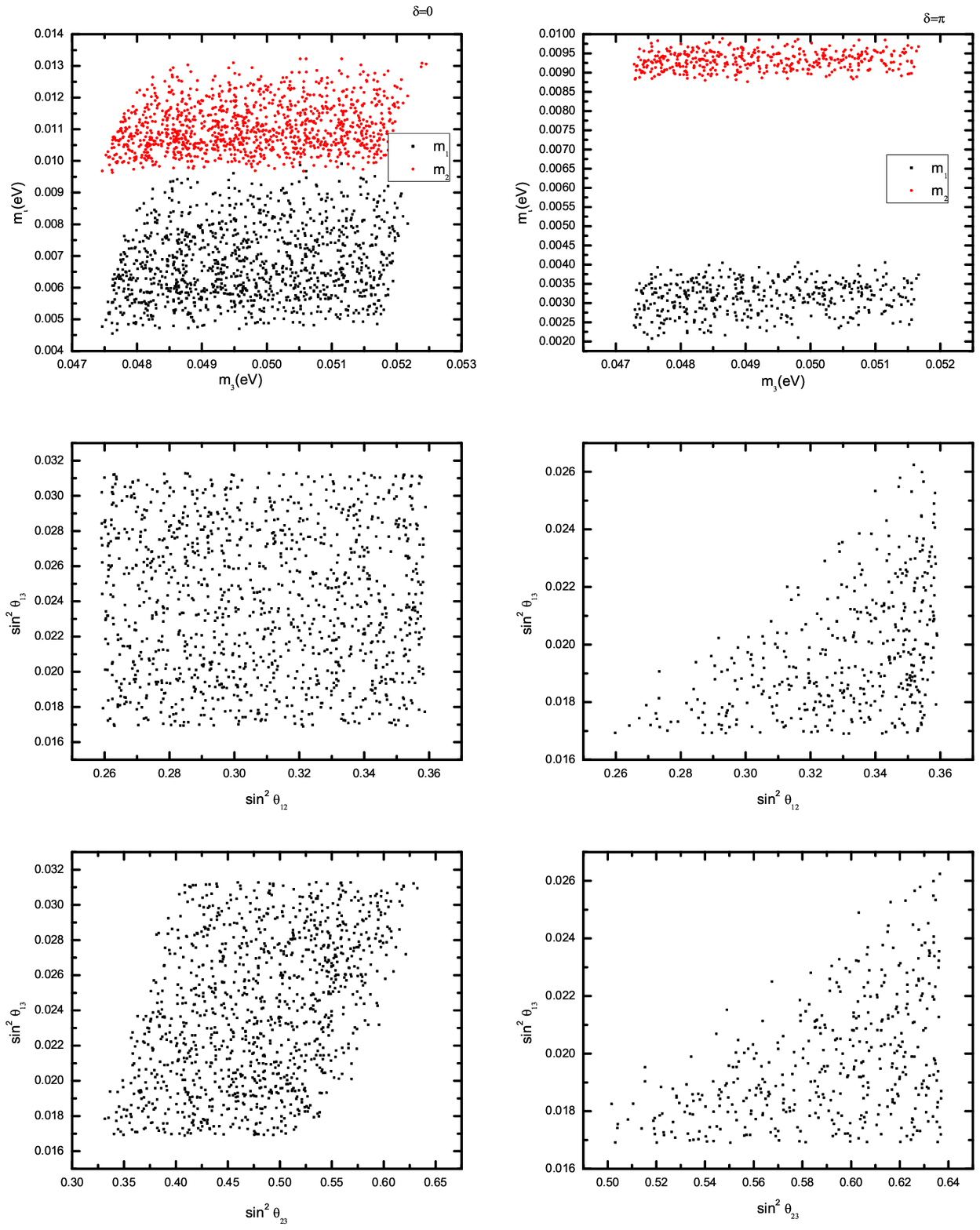,bbllx=3cm,bblly=13cm,bburx=18cm,bbury=28cm,%
width=13cm,height=13cm,angle=0,clip=0}} \vspace{3.5cm}
\caption{${\bf Pattern~A^{}_2}$ of $M^{}_\nu$: The allowed regions
of neutrino masses $(m^{}_1, m^{}_2, m^{}_3)$ and neutrino mixing
angles $(\sin^2 \theta^{}_{12}, \sin^2 \theta^{}_{23}, \sin^2
\theta^{}_{13})$, where the left column is for $\delta=0$ while the
right column for $\delta=\pi$. The global-fit data of two neutrino
mass-squared differences $(\delta m^2, \Delta m^2)$ and three
neutrino mixing angles $(\theta^{}_{12}, \theta^{}_{23},
\theta^{}_{13})$ at the $3\sigma$ level have been input.}
\end{figure}
%%%%%%%%%%%%%%%%%%%%%%%%%%%%%%%%%%%%%%%%%%%%

%%%%%%%%%%%%%%%%%% Fig. 3 %%%%%%%%%%%%%%%
\begin{figure}
\centerline{\psfig{file=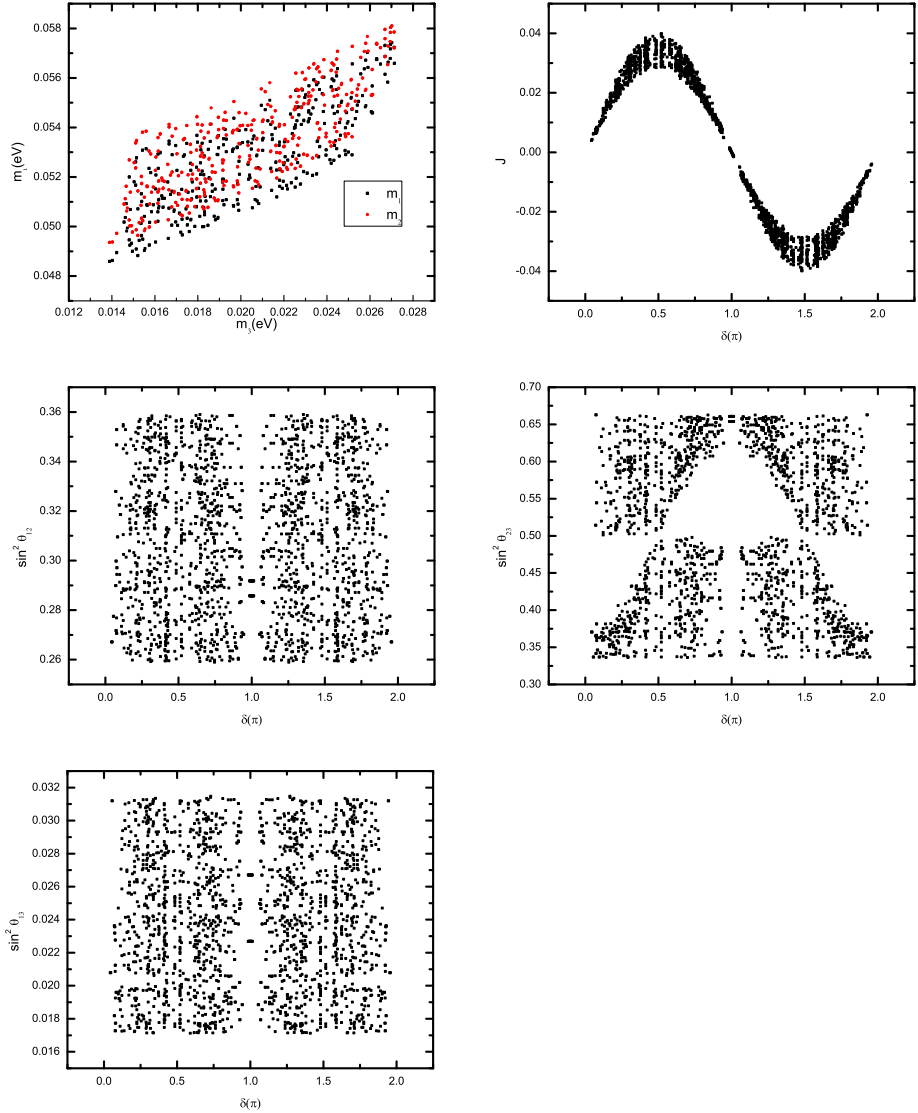,bbllx=3cm,bblly=13cm,bburx=18cm,bbury=28cm,%
width=13cm,height=13cm,angle=0,clip=0}} \vspace{3.5cm}
\caption{${\bf Pattern~C}$ of $M^{}_\nu$: The allowed regions of
neutrino masses $(m^{}_1, m^{}_2, m^{}_3)$, neutrino mixing angles
$(\sin^2 \theta^{}_{12}, \sin^2 \theta^{}_{23}, \sin^2
\theta^{}_{13})$, and the Jarlskog invariant versus the CP-violating
phase $(J, \delta)$. The global-fit data of two neutrino
mass-squared differences $(\delta m^2, \Delta m^2)$ and three
neutrino mixing angles $(\theta^{}_{12}, \theta^{}_{23},
\theta^{}_{13})$ at the $3\sigma$ level have been input, while the
CP-violating phase $\delta$ is allowed to freely vary in $[0,
2\pi)$.}
\end{figure}
%%%%%%%%%%%%%%%%%%%%%%%%%%%%%%%%%%%%%%%%%%%

\end{document}